\RequirePackage[2020-02-02]{latexrelease}
\documentclass[twocolumn]{aastex631}

\received{---}
\revised{---}
\accepted{---}
\submitjournal{ApJ}

\shorttitle{A proto-cluster of massive quiescent galaxies at $z=4$}
\shortauthors{Tanaka et al.}

\begin{document}

\title{A proto-cluster of massive quiescent galaxies at $z=4$}

\correspondingauthor{Masayuki Tanaka}
\email{masayuki.tanaka@nao.ac.jp}

\author{Masayuki Tanaka}
\affil{National Astronomical Observatory of Japan, 2-21-1 Osawa, Mitaka, Tokyo 181-8588, Japan}
\affil{Graduate Institute for Advanced Studies, SOKENDAI, 2-21-1 Osawa, Mitaka, Tokyo 181-8588, Japan}

\author{Masato Onodera}
\affil{Subaru Telescope, National Astronomical Observatory of Japan, 650 N Aohoku Pl, Hilo, HI96720}
\affil{Graduate Institute for Advanced Studies, SOKENDAI, 2-21-1 Osawa, Mitaka, Tokyo 181-8588, Japan}

\author{Rhythm Shimakawa}
\affil{Waseda Institute for Advanced Study (WIAS), Waseda University, Nishi Waseda, Shinjuku, Tokyo 169-0051, Japan}
\affil{Center for Data Science, Waseda University, 1-6-1, Nishi-Waseda, Shinjuku, Tokyo 169-0051, Japan}

\author{Kei Ito}
\affil{Department of Astronomy, School of Science, The University of Tokyo, 7-3-1, Hongo, Bunkyo-ku, Tokyo, 113-0033, Japan}

\author{Takumi Kakimoto}
\affil{Graduate Institute for Advanced Studies, SOKENDAI, 2-21-1 Osawa, Mitaka, Tokyo 181-8588, Japan}

\author{Mariko Kubo}
\affil{Astronomical Institute, Tohoku University, 6-3, Aramaki, Aoba-ku, Sendai, Miyagi, 980-8578, Japan}

\author{Takahiro Morishita}
\affil{IPAC, California Institute of Technology, MC 314-6, 1200 E. California Boulevard, Pasadena, CA 91125, USA}

\author{Sune Toft}
\affil{Cosmic Dawn Center (DAWN)}
\affil{Niels Bohr Institute, University of Copenhagen, Jagtvej 128, DK-2200 Copenhagen, Denmark}

\author{Francesco Valentino}
\affil{European Southern Observatory, Karl-Schwarzschild-Str. 2, D-85748 Garching bei Munchen, Germany}
\affil{Cosmic Dawn Center (DAWN)}

\author{Po-Feng Wu}
\affil{Graduate Institute of Astrophysics and Department of Physics, National Taiwan University, No.1, Sec. 4 Roosevelt Road, Taipei 10617, Taiwan}
\affil{Department of Physics and Center for Theoretical Physics, National Taiwan University, No.1, Sec. 4 Roosevelt Road, Taipei 10617, Taiwan}
\affil{Physics Division, National Center for Theoretical Sciences, No.1, Sec. 4 Roosevelt Road, Taipei 10617, Taiwan}

\begin{abstract}
  We report on discovery of a concentration of massive quiescent galaxies located at $z=4$.
  The concentration is first identified using high-quality photometric redshifts based on deep,
  mutli-band data in Subaru/XMM-Newton Deep Field.  Follow-up near-infrared spectroscopic
  observations with MOSFIRE on Keck confirm a massive ($\sim10^{11}\rm\ M_\odot$) quiescent galaxy at $z=3.99$.
  Our spectral energy distribution (SED) analyses reveal that the galaxy experienced an episode of starburst about 500~Myr prior to the observed epoch,
  followed by rapid quenching.  As its spectrum is sufficiently good to measure the stellar
  velocity dispersion, we infer its dynamical mass and find that it is consistent with its stellar mass.
  The galaxy is surrounded by 4 massive ($>10^{10}\rm\ M_\odot$) quiescent galaxies on a $\sim1$ physical Mpc scale,
  all of which are consistent with
  being located at the same redshift based on high-accuracy spectro-photometric redshifts.
  This is likely a (proto-)cluster dominated by quiescent galaxies, the first of the kind reported
  at such a high redshift as $z=4$.
  Interestingly, it is in a large-scale structure revealed by spectroscopic redshifts from VANDELS. Furthermore,
  it exhibits a red sequence, adding further support to the physical
  concentration of the galaxies. We find no such concentration in the Illustris-TNG300 simulation;
  it may be that the cluster is such a rare system that the simulation box is not sufficiently large
  to reproduce it.
  The total halo mass of the quiescent galaxies is $\sim10^{13}~\rm M_\odot$, suggesting that
  they form a group-sized halo once they collapse together.
  We discuss implications of our findings for the quenching physics and conclude
  with future prospects.
\end{abstract}


\section{Introduction}
\label{sec:intro}

Galaxies in the local Universe exhibit a large diversity in their morphologies as characterized by the Hubble Sequence.
Most massive galaxies tend to be elliptical and they show little sign of on-going star formation activities.
They form a tight sequence on a color-magnitude diagram (red sequence; \citealt{bower92}) and the tightness indicates that they are a very
homogeneous population.  Their spectra indeed show that they are all dominated by old stars formed long time ago.
Furthermore, detailed absorption line analyses suggest that they formed in an intense burst of star formation occurred in
a short timescale, followed by passive evolution (i.e., simple aging without formation of new stars; \citealt{thomas10}). 
This is an intriguing observational fact because these galaxies are massive, and we expect massive galaxies to appear
only at low redshifts in the current paradigm of hierarchical galaxy formation.
These observations pose three major questions:

\begin{itemize}
\item What physical process(es) drives the intense starburst in the early Universe?
\item What causes the rapid shutdown of the starburst activity?
\item What physical process(es) prevents subsequent formation of new stars for $>$10~Gyr?
\end{itemize}

\noindent
These questions are indeed among the most outstanding questions in the field of galaxy formation and evolution for decades.
This paper aims to address the first two questions.

An interesting observational approach to these questions is to go back in time and study progenitors of
nearby massive elliptical galaxies in the early Universe.  A lot of efforts have been put in the search for distant quiescent
galaxies (i.e., galaxies with no star formation) and quiescent galaxies have been identified up
to $z\sim4$ and beyond \citep{schreiber18,tanaka19,valentino20,kakimoto23} thanks to the advent of sensitive near-infrared spectrographs on the ground.  These galaxies
are post-starburst galaxies; their star formation histories (SFHs) inferred from extensive fitting of
their SEDs with stellar population synthesis models indicate that they all underwent an intense starburst
in a recent past and then experienced rapid quenching.  We are getting very close to the formation epoch
of the nearby elliptical galaxies.  However, as we show below, $z\sim4$ is the spectroscopic sensitivity limit
from the ground and it is extremely challenging to go further.

JWST has already started to identify quiescent galaxies at $z \gtrsim 3$ from its very first observations
\citep{carnall22,carnall23,marchesini23,valentino23},
and it will surely revolutionize our understanding of the formation of massive quiescent galaxies.
Due to its small field coverage, however, one key aspect will be missed in blank field JWST observations: environment.
In the local Universe, we know galaxy properties are strongly dependent on their surrounding environment \citep{dressler80}.
Quiescent early-type galaxies dominate the dense cluster core, while star-forming spiral galaxies
are the main population in the low-density field.  Also, we expect that galaxies form earlier in
higher density peaks of the density fluctuation in the Universe; galaxy formation is accelerated in
cluster regions.  However, massive clusters are rare objects in the Universe especially at high redshifts.
Thus, a wide-field observation is required to understand the role of environment in the formation of
massive quiescent galaxies in the distant Universe.

There have been a lot of efforts to search for (proto-)clusters using
star-forming galaxies as a tracer (e.g., \citealt{miller18,oteo18,toshikawa18}).
However, much less attention has been 
paid to quiescent galaxies due primarily to difficulties in their identifications at high redshifts;
they are faint and rare objects ($\sim10^{-6}\rm\  Mpc^{-3}$; \citealt{schreiber18,valentino20,marsan22,valentino23}),
requiring deep imaging data over a wide area.
Deep multi-wavelength data in fields like the Cosmic Evolution Survey (COSMOS) field make it possible to search for them in the early Universe.
For example, \citet{strazzullo15} searched for concentrations of quiescent galaxies in the COSMOS field out to $z\sim2.5$.
Recently, \citet{ito23a} successfully confirmed a concentration of
massive quiescent galaxies with deep near-IR spectroscopy at $z=2.7$.
\citet{kubo15,kubo21} identified a massive quiescent galaxy in a proto-cluster at $z=3.1$.
The currently known high-$z$ proto-clusters with quiescent galaxies are still only a few and are limited to $z\lesssim3$.
In this paper, we focus on quiescent galaxies at $z=4$ based on deep spectroscopic observations and discuss an early structure
populated by these quiescent galaxies to address the role of environment.

The paper is structured as follows.
We describe our spectroscopic observations in Section 2, followed by spectroscopic analyses of
a confirmed quiescent galaxy at $z=3.99$ in Section 3. We then focus on a concentration of
several quiescent galaxies at the same redshift using joint photometric and spectroscopic fits to their observed
spectral energy distributions (SEDs) in Section 4. Section 5
summarizes physical properties of the concentration such as the Butcher-Oemler effect \citep{butcher84} at $z=4$. Finally,
we discuss our results in the context of massive galaxy formation and conclude the paper in Section 6. 
We adopt $\rm H_0=70\ km\ s^{-1}\ Mpc^{-1}$, $\rm \Omega_M=0.3$, and $\rm \Omega_\Lambda=0.7$.  Magnitudes are in the AB system
\citep{oke83}.

\section{Observation}
\label{sec:observation}

\label{sec:target_selection}

We focus on the Subaru/XMM-Newton Deep Survey (SDXS), where very deep, multi-wavelength imaging data
are available over a wide area.  We have constructed a multi-band photometric catalog as summarized
in \citet{kubo18}.  This is the same catalog that \citet{tanaka19} and \citet{valentino20} are
based on.  We apply the photometric redshift code from \citet{tanaka15} to the catalog to infer
photometric redshifts as well as physical properties of galaxies such as stellar mass and star
formation rate (SFR).
In brief, the code uses models from \citet{bruzual03} over a wide parameter range with template
error functions to account for systematic uncertainties in the models.  We adopt the \citet{chabrier03}
initial mass function.

The code is calibrated to relatively low-redshift objects with spectroscopy available
at that time.  Since then, VANDELS \citep{pentericci18,garilli21} has carried out an extensive spectroscopic campaign focused
on high-redshift galaxies in the field and released their catalog to the public.
We compare our photometric redshifts with secure spectroscopic redshifts from VANDELS DR4 with zflag$=3$ or $4$
in Fig.~1.
We note that the VANDELS targets are selected on the basis of photometric redshifts. There are a few different
types of targets, but we are primarily using quiescent galaxies at $1<z<2$ and star-forming galaxies at $2.5<z<5$ here.

We emphasize that Fig.~1 is entirely a blind comparison; VANDELS spec-$z$’s were not available when we computed our photometric
redshifts.  As can be seen, the correspondence is excellent:
an outlier rate defined as the fraction of objects with $|\delta z|/(1+z)>0.15$ \citep{tanaka18} is 4.8\% and the dispersion
defined as $\sigma((\delta z)/(1+z))$ is 0.036.
The outliers scatter down to low redshifts, but most of them are consistent with spec-$z$'s within the uncertainties.

As VANDELS does not include quiescent galaxies at high redshifts, we supplement it with high-$z$
quiescent galaxies from the literature (\citealt{schreiber18,tanaka19}) as well as from this paper and show them
as the red circles in Fig.~1. While the redshift distribution of the quiescent galaxies is very sparse,
the photo-$z$ accuracy for the quiescent galaxies at $z\sim4$ is similar to that of star-forming galaxies from VANDELS.
The figure demonstrates that our photometric redshifts are accurate even for quiescent galaxies at high redshifts.

\begin{figure}
\centering
\includegraphics[width=80mm]{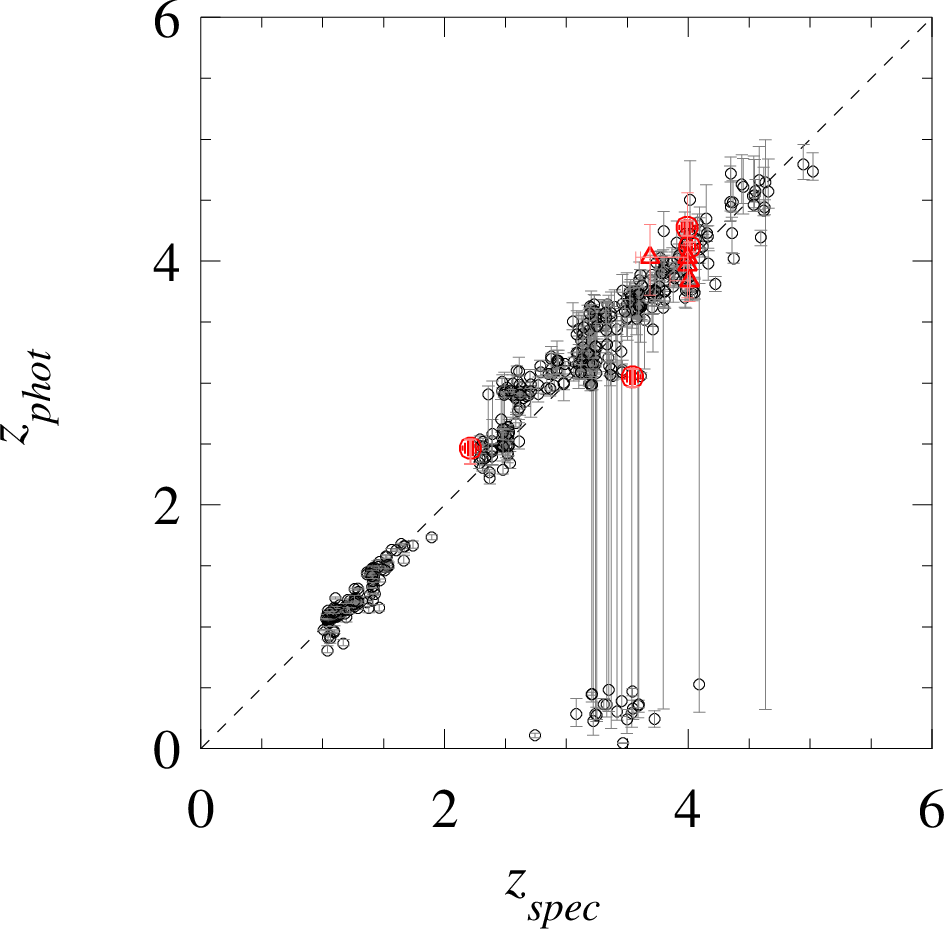}
\caption{
  Photometric redshifts plotted against the secure spectroscopic redshifts from VANDELS DR4.
  The error bars indicate the $68\%$ confidence intervals.
  The red circles are spectroscopically confirmed quiescent galaxies from \citet{schreiber18}, \citet{tanaka19}, and this paper.
  We differentiate spectrophotometric galaxies (see Section 4) from
  the secure spectroscopic redshifts with the open triangles.
  The dashed line shows the $z_{phot}=z_{spec}$ relation.
}
\label{fig:photoz_vs_specz}
\end{figure}

Motivated by the good accuracy, we search for an over-density of massive quiescent galaxies at
high redshifts.
We search for over-density regions by selecting quiescent galaxies in a redshift slice of $\delta z\pm0.3$ and
evaluate their over-density significance as a function of position.
This redshift slice roughly corresponds to $\pm1.5\sigma(z_{phot})$.
Quiescent galaxies are defined
as those with $1\sigma$ upper limit of specific SFR (sSFR) being smaller than $10^{-9.5}\rm Gyr^{-1}$,
which corresponds to $\sim1$ dex below the star formation main sequence, as we have
done in our previous papers \citep{kubo18,valentino20,ito22,ito23a}.   We identify a significant
over-density of quiescent galaxies at $z\sim4$ in SXDS as shown in Fig.~2; based on the kernel
density estimate with Gaussian width of 2.4 arcmin (1 physical Mpc at $z=4$), the over-density
significance is $\sim20\sigma$. 
\citet{tanaka19} reported the discovery of the $z=4.01$ galaxy in the same field and at the same redshift
as the over-density region.  This is not a coincidence as we discuss below.  The over-density region
is approximately 5 arcmin ($\sim2$ physical Mpc) on a side.

\begin{figure}
\centering
\includegraphics[width=80mm]{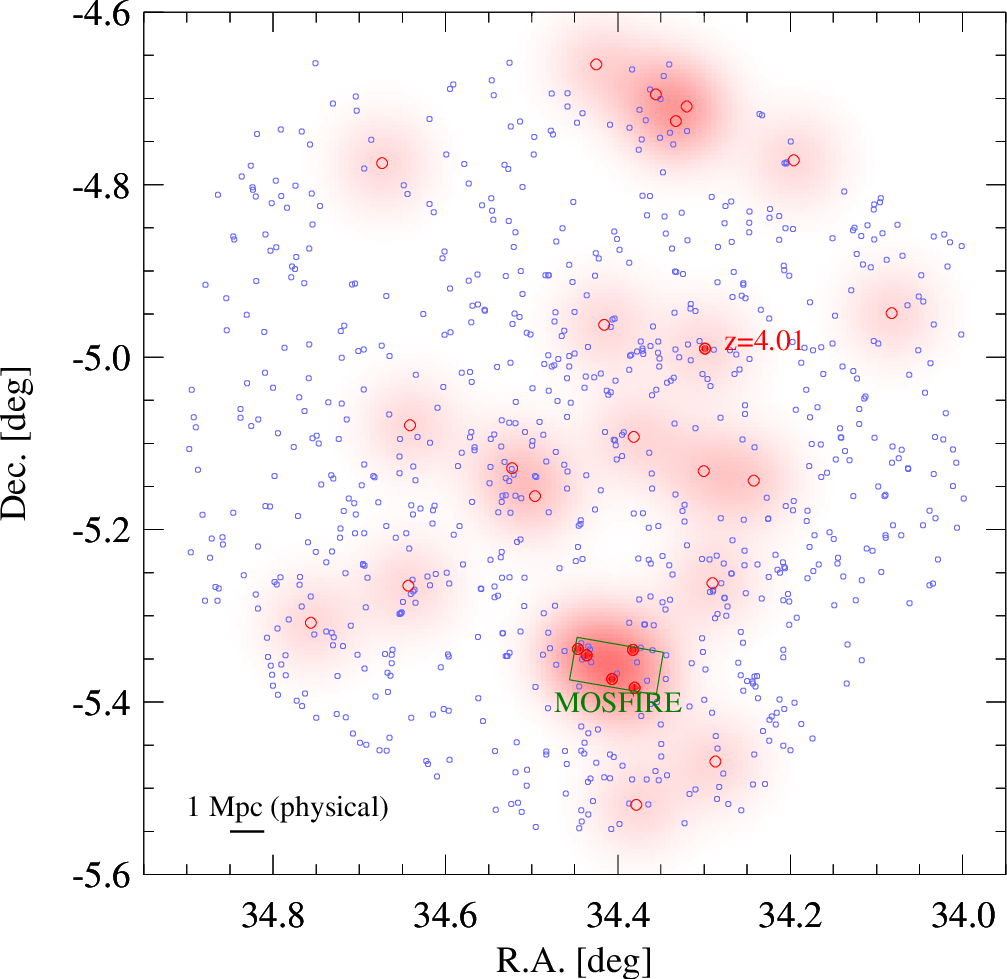}
\caption{
  Distribution of massive ($M_*>10^{10}\rm M_\odot$) galaxies at $z\sim4$ in SXDS.
  The open red/blue circles are photo-$z$ selected quiescent/star-forming galaxies at $3.7<z_{phot}<4.3$.
  The color contours show the density of quiescent galaxies smoothed over 1 Mpc (physical).
  The filled red points are galaxies in the over-density region and are the focus of the paper.
  There is a spectroscopically confirmed $z=4.01$ galaxy in the north of the over-density region \citep{tanaka19}.
  The green square is the field field of view of our MOSFIRE observation.
  The bar on the bottom-left indicates 1 Mpc (physical) at $z=4$.
}
\label{fig:z4_distrib}
\end{figure}

\subsection{Observation and Data Reduction}
\label{sec:data_reduction}

We carried out a follow-up spectroscopic observation of the region with MOSFIRE \citep{mclean10}
on the Keck I telescope on Nov 29 - Dec 2, 2020.
As the over-density region is compact, we could observe all the quiescent targets
in a single MOSFIRE field of view. We observed in the $K$-band and applied the ABBA
dither along the 0.7 arcsec width slit with an individual exposure time of 3 min
and a nod amplitude of $\pm 1.25$ arcsec. The conditions were clear with slightly
variable seeing between 0.5 and 1.2 arcsec with a median of 0.74 arcsec.
The data were reduced with the MOSFIRE DRP 2018 release with custom scripts \citep{kakimoto23}.
We detail our data processing in Appendix A.
The total integration was $\sim16$ hours.

We observed 5 quiescent galaxy candidates as summarized in Table 1.
Only one of them is sufficiently bright to measure its redshift from absorption features and its
spectrum is shown in Fig.~3. The spectrum is rather noisy (the median S/N per pixel
is 1.8), but we can still identify multiple Balmer absorption lines, which yield $z_{spec}=3.987\pm0.001$.
The continuum is relatively flat overall, but there is a hint of the 4000\AA\ break on the blue end,
which indicates an evolved stellar population. We will discuss this object in detail in the next Section.
We could not measure redshifts for the reminder of the quiescent galaxy candidates observed
in the same mask as they are fainter
(see Table 1).
We attempt to constrain their redshifts
using both the spectra and photometry and discuss their properties in Sections~4
and 5.

\begin{deluxetable*}{lcccccc}
  \tablecaption{
    Coordinates and redshifts of the quiescent galaxies
  }
  \tablehead{
    \colhead{ID} & \colhead{R.A.} & \colhead{Dec.} & \colhead{$K$-band mag.} & \colhead{$z_{phot}$} & \colhead{$z_{spec}$} & \colhead{$z_{specphot}$}\\
  }
  \startdata
  SXDS2\_19838 & $02^h 17^m 31.8$ & $-05^\circ 20' 22''.6$ & 22.72 & $4.28^{+0.06}_{-0.07}$ & $3.99$ & $3.985^{+0.003}_{-0.003}$\\
  SXDS2\_15659 & $02^h 17^m 31.4$ & $-05^\circ 23' 00''.0$ & 23.46 & $3.97^{+0.16}_{-0.19}$ & $-$ &  $3.993^{+0.005}_{-0.030}$\\
  SXDS2\_16609 & $02^h 17^m 37.6$ & $-05^\circ 22' 24''.4$ & 23.99 & $4.03^{+0.27}_{-0.31}$ & $-$ &  $3.687^{+0.271}_{-0.081}$\\
  SXDS2\_19229 & $02^h 17^m 44.6$ & $-05^\circ 20' 44''.6$ & 23.74 & $3.84^{+0.24}_{-0.17}$ & $-$ & $4.009^{+0.009}_{-0.158}$\\
  SXDS2\_19997 & $02^h 17^m 47.1$ & $-05^\circ 20' 19''.2$ & 23.67 & $3.94^{+0.53}_{-0.33}$ & $-$ &  $3.993^{+0.005}_{-0.423}$\\
  \hline
  SXDS5\_27434 & $02^h 17^m 11.7$ & $-04^\circ 59' 23''.5$ & 21.89 & $4.12^{+0.04}_{-0.04}$ & $4.01$ & $4.005^{+0.003}_{-0.003}$\\
  \enddata
  \tablecomments{
    $z_{phot}$ is based on the broad-band photometry only, while $z_{specphot}$ uses the spectra as well (Section 4).
    The last object (SXDS5\_27434) is from \citet{tanaka19} for reference.
  }
  \label{tab:coords}
\end{deluxetable*}

\begin{figure*}
\centering
\includegraphics[width=170mm]{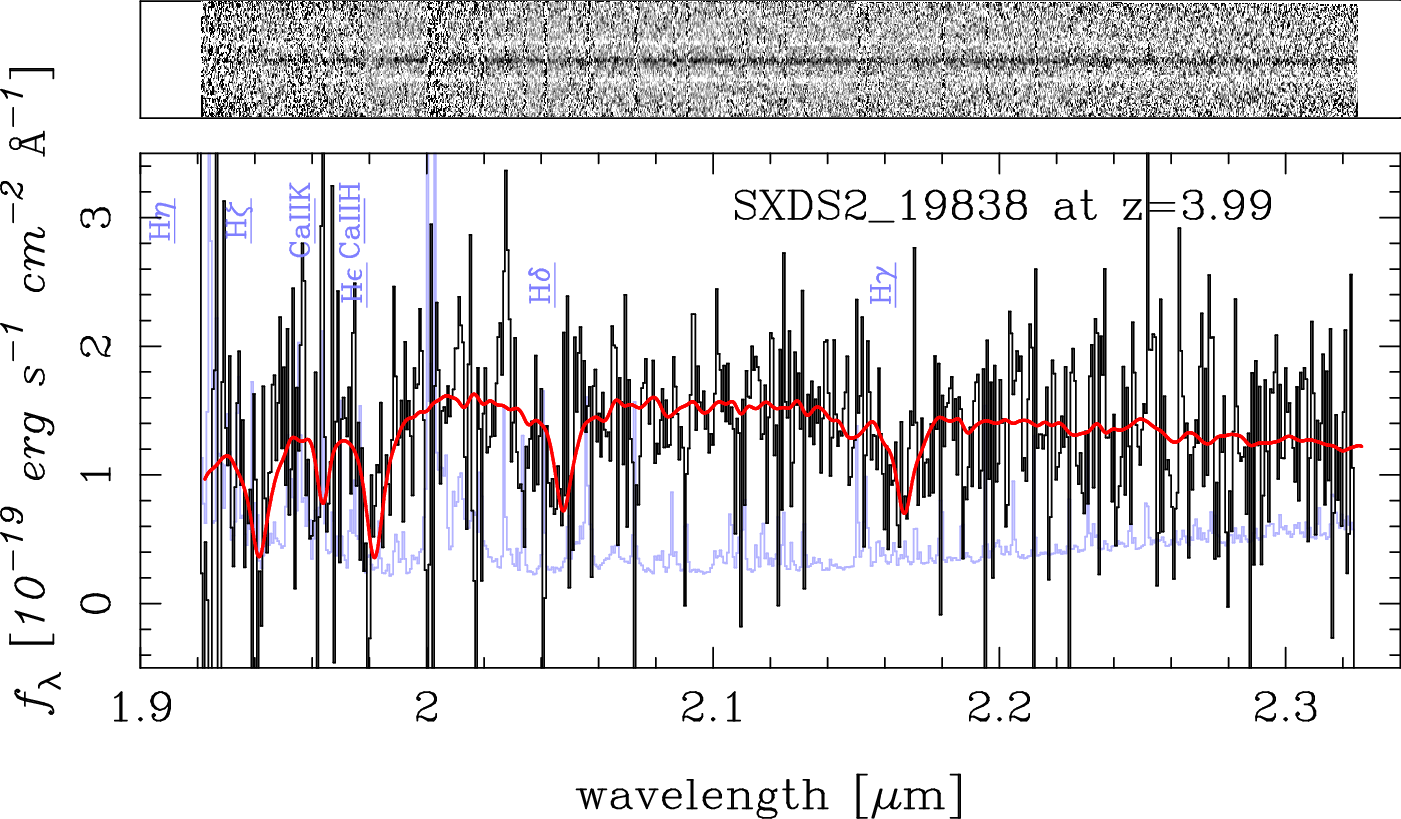}
\caption{
  The top and bottom panels show the 2d and 1d spectrum, respectively.  In the bottom panel, the black and
  blue spectra are the observed and noise spectra binned by 3 pixels.  Some of the most prominent absorption features are
  indicated.  The red spectrum is the best-fit {\tt ppxf} model (see Section~3.2 for details).
}
\label{fig:spectrum}
\end{figure*}

\section{Detailed Properties of the Confirmed $z=4$ Galaxies}
\label{sec:spectral_analysis}

In this Section, we examine properties of the $z=3.99$ galaxy in detail and compare them with
those of the $z=4.01$ galaxy presented in \citet{tanaka19} and \citet{valentino20}.
We closely follow the analysis of \citet{tanaka19} in this Section.
We measure its physical size (Section 3.1), and discuss the dynamical properties
(Section 3.2).
We then infer its star formation histories (Section 3.3).
As we show below, the two galaxies are both massive galaxies with
suppressed SFRs, but there are interesting differences. The $z=3.99$ galaxy is in
an over-density region (Section~5), while the $z=4.01$ galaxy
is not. We compare their properties with their environmental differences in mind.

\subsection{Size-Mass Relation}
\label{sec:size_mass_relation}

We first focus on the size vs. stellar mass relation.
Galaxies are known to exhibit a tight size-mass relation in the sense that more massive
galaxies are physically larger \citep{shen03}.  Interestingly, quiescent galaxies tend to be more compact
than star-forming galaxies of similar stellar mass.  This relation evolves with redshift;
galaxies at fixed stellar mass are smaller at higher redshifts (e.g., \citealt{vanderwel16}).
The evolution has been tracked up to $z\sim4$ and beyond \citep{kubo18,ito23a} and massive quiescent
galaxies there are extremely compact and dense.  We briefly address this size-mass relation at $z\sim4$.

We do not have high-resolution imaging data of the $z=3.99$ galaxy, but as shown in \citet{tanaka19},
the ground-based high-quality images deliver sizes consistent with those from {\it Hubble Space Telescope}.
Furthermore, the AO-assisted $K_s$-band imaging of the $z=4.01$ galaxy presented in \citet{ito23b}
is consistent with that measured from the seeing-limited imaging data from the Hyper Suprime-Cam Subaru
Strategic Program (HSC-SSP; \citealt{aihara18a,aihara22}).
We thus adopt the same approach here to measure the size of the $z=3.99$ galaxy. That is
to use deep optical data from HSC-SSP.
We fit the $i$-band image, in which the galaxy is detected at $S/N\sim20$, with {\tt galfit} \citep{peng02,peng10}
to measure its effective radius assuming the S\'{e}rsic profile.
The coadd PSF from the HSC pipeline \citep{bosch18} is used as an input PSF image (FWHM$=0.72$ arcsec).
We run a Monte-Carlo simulation by perturbing the initial guess on the centroid, position angle, effective radius, and brightness.
In each run, the S\'{e}rsic index is fixed to a randomly drawn value between 0.5 and 4.
The fit with the smallest $\chi^2$ is taken as the best estimate and $\Delta \chi^2<1$ as the uncertainty.
We obtain $r_{eff}=1.04\pm0.43$ kpc.
Note that we use the effective radius along the semi-major axis throughout the paper.
We use stellar mass from the joint photometric and spectroscopic fitting from Section 4.

Fig.~4 shows the size vs. stellar mass relation including results from the literature.
The sizes of our $z\sim4$ galaxies ($z=3.99$ and $z=4.01$) are clearly smaller than quiescent galaxies
at $z\lesssim2$, suggesting a significant size evolution.
Note that the size of the $z=4.01$ galaxy is from \citet{ito23b}.
Our measurements are
in line with the literature results at $z\sim3.5$; massive quiescent galaxies at
these redshifts are smaller than $\sim1$ kpc, despite their large mass.  This
makes them a very dense system; as discussed in \citet{kubo18}, their surface
stellar mass densities are comparable to those of globular clusters.  One
possible formation mechanisms for such a dense system is nuclear starburst.
Their star formation histories (Section 3.3) indeed suggest that they experienced
an intense burst of star formation in a recent past.


\begin{figure}
\centering
\includegraphics[width=80mm]{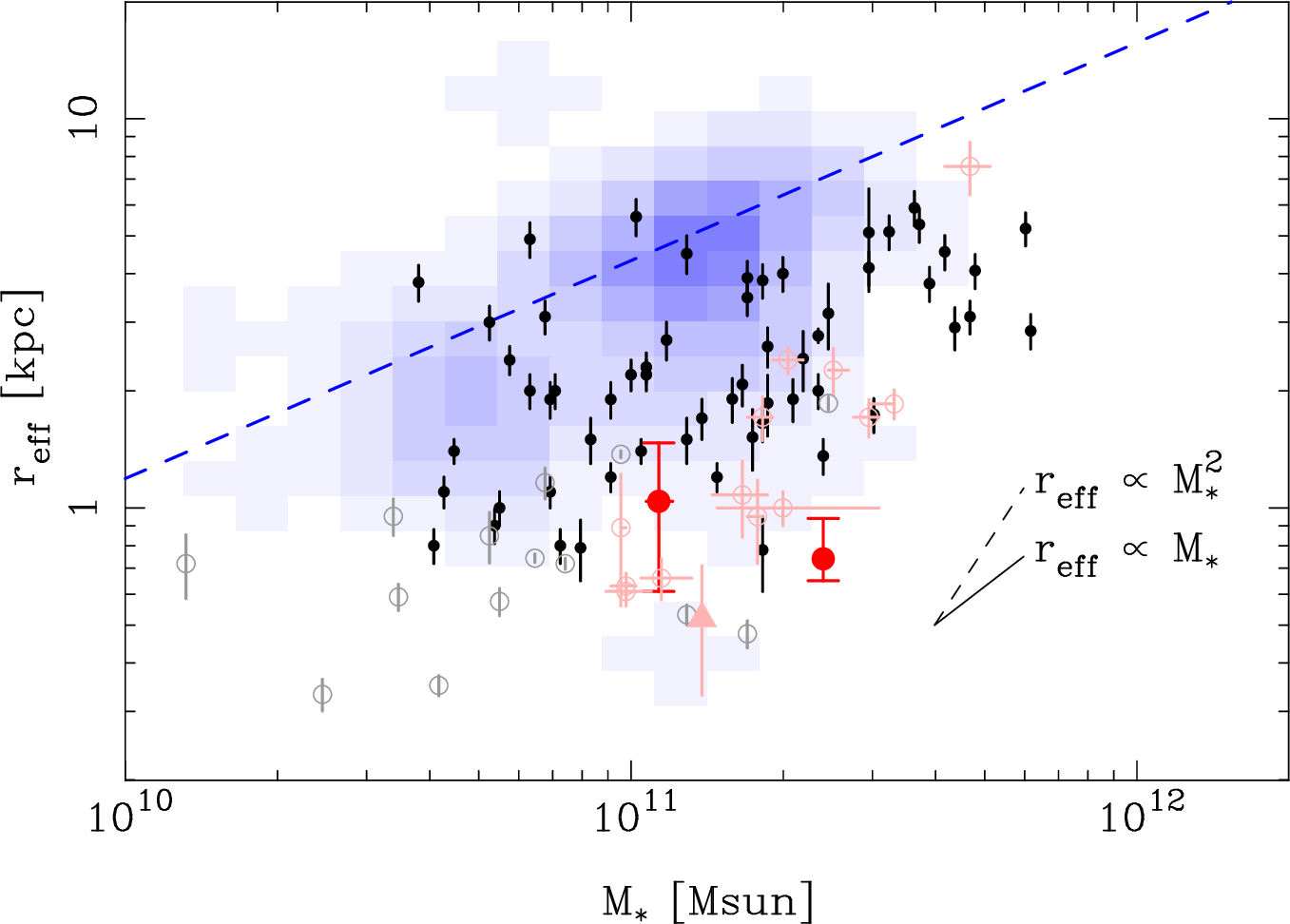}
\caption{
  Effective radius plotted against stellar mass. The dashed line is the $z=0$ relation
  from \citet{shen03}, and the blue shades are quiescent galaxies at $z\sim0.7$ from the LEGA-C survey
  \citep{vanderwel16,straatman18}.  Quiescent galaxies at $z\sim2$ are collected from the literature
  \citep{vandokkum09,onodera12,toft12,bezanson13,vandesande13,belli17,stockmann20}
  and are indicated as the black points.
  Another collection from the literature of $z\sim3.5$ galaxies are shown in pink
  \citep{saracco20,esdaile21,forrest22}.
  The stacked object of \citet{kubo18} is shown as the filled triangle.
  The gray symbols are the JWST measurements of quiescent galaxies at $z>3$ from \citet{ito23b}.
  Note that they are photo-$z$ selected galaxies (i.e., no spectroscopic confirmation yet).
  Our $z=4$ objects are in red.  The error bars show the statistical uncertainties.
  The solid and dotted lines in the bottom right corner show evolutionary tracks with
  $r_{eff}\propto M_*$ and $r_{eff}\propto M_*^2$, respectively. They represent the major and minor merger tracks
  \citep{bezanson09,naab09}.
}
\label{fig:re_vs_mstar}
\end{figure}

\subsection{Dynamical Properties}
\label{sec:dynamical_properties}

We move on to discuss dynamical properties of the $z=4$ galaxies with a focus
on the stellar velocity dispersion here.
We do not go very far in this dynamical analysis because the velocity dispersion is not easy to
estimate robustly in young galaxies like ours and the galaxies may be rotating \citep{toft17,newman18,deugenio23},
which makes the interpretation difficult.

The observed MOSFIRE spectrum of the $z=3.99$ galaxy is fit with {\tt ppxf} \citep{cappellari04,cappellari17}.
Simple stellar population (SSP) models from \citet{vazdekis10} are used in the fit.
Following \citet{tanaka19}, we exclude models older than the age of the universe at the redshift of the object
as well as those with subsolar metallicities as we deal with a massive galaxy.
We use an additive correction function of order 1 and no multiplicative correction.
Our results are not sensitive to the choices here.  We run {\tt ppxf} by perturbing the each pixel of
the spectrum by its uncertainty and repeat the fit.  We measure $\sigma=305\pm103\ \rm km\ s^{-1}$.

The correlation between the stellar velocity dispersion and stellar mass is plotted in Fig.~5.
Again, we include the $z=4.01$ galaxy ($\rm M_* \sim2\times10^{11}M_\odot$).  The $z<1$ relations are 
fully consistent with each other, while $z\sim2$ galaxies seem to show a tail towards
larger $\sigma$.  Galaxies at $z\sim3.5-4$ also exhibit slightly larger $\sigma$ on average compared to the $z<1$ relations.
The scatter is large, but this may indicate a mild evolution of the stellar velocity dispersion at high redshifts.
However, as mentioned earlier, the interpretation is tricky as they may be rotating disk galaxies \citep{toft17,newman18}.
It is now feasible to measure Sersic indices of high redshift galaxies with JWST (e.g., \citealt{ito23b}),
and a joint analysis of morphology and stellar dynamics will be an interesting avenue.

\begin{figure}
\centering
\includegraphics[width=80mm]{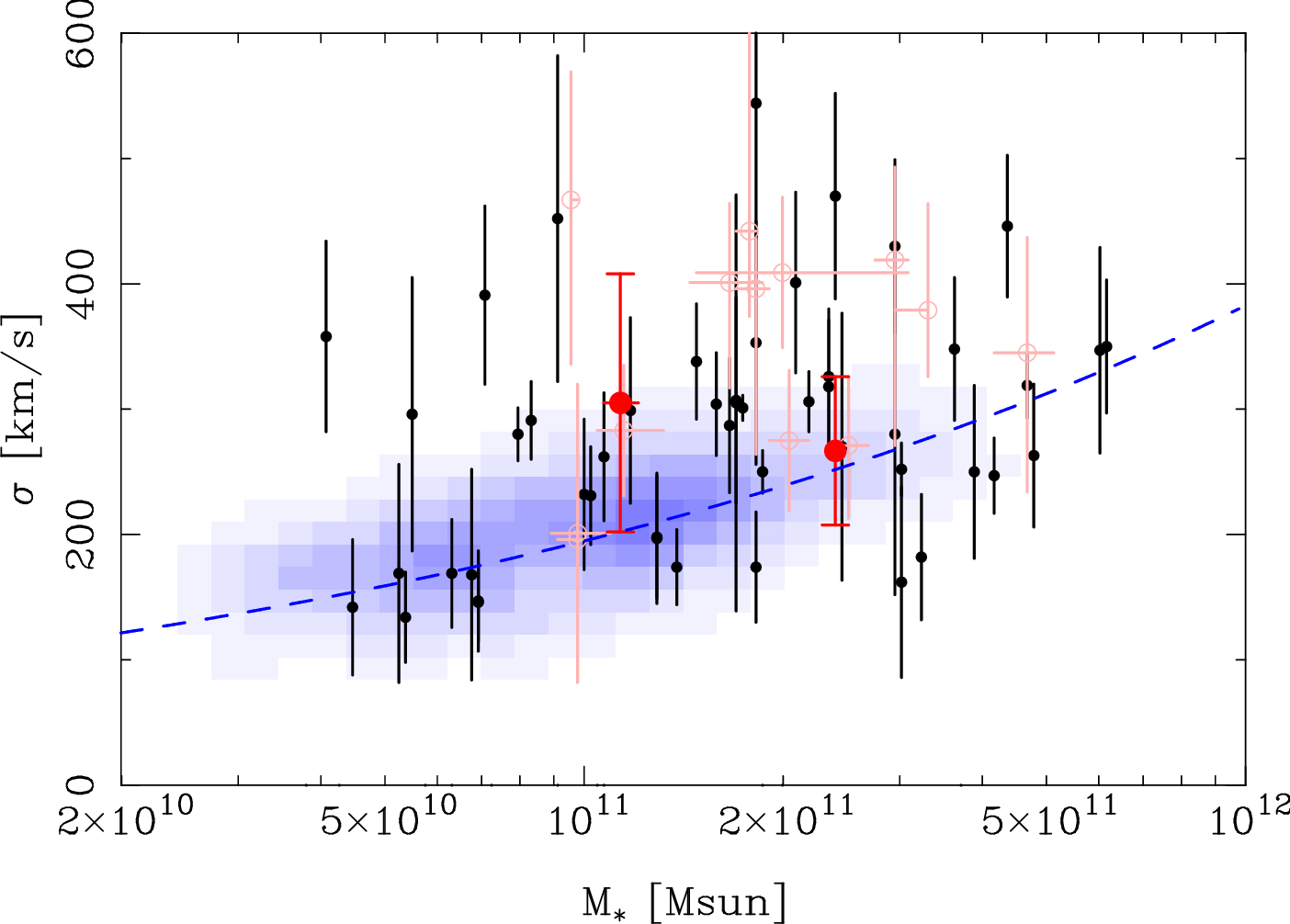}
\caption{
  Stellar velocity dispersion is plotted against stellar mass.
  The blue dashed curve is for $z=0$ galaxies and is from \citet{zahid16}.
  The meanings of the other symbols are the same as in Fig.~4.
}
\label{fig:sigma_vs_mstar}
\end{figure}

\subsection{Star Formation Histories}
\label{sec:star_formation_histories}

The spectral fitting in the previous subsection fits the observed spectrum with combination of
SSP models with different ages.  We thus have relative weights between different ages for each fit.
We use this information to infer the star formation histories of the $z=4$ galaxies.
The age grid is discrete, and we apply a Gaussian kernel to make it contiguous.
The inferred star formation histories (SFHs) are shown in Fig.~6.
We note that the SFHs here are based on the spectrum only. The photometric data may add further constraints,
and we leave in-depth SFH analyses using both the spectra and photometry for all
the quiescent galaxies (see the next Section for the fainter quiescent galaxies) for future work.

There are both similarities and differences between the two SFHs of the $z=3.99$ and $4.01$ galaxies;
these two galaxies experienced a starburst, followed by a rapid
decrease in their SF activities.  This SFH is very similar to those
observed in other massive quiescent galaxies at high redshifts \citep{schreiber18,valentino20}
and even those inferred from the local universe (e.g., \citealt{thomas10}).
It is interesting that the $z=3.99$ galaxy experienced the burst earlier than the $z=4.01$.
In other words, the $z=3.99$ galaxy is older. The $z=3.99$ galaxy is in a dense region as we discuss below,
while the $z=4.01$ galaxy is isolated. We cannot draw any conclusion with just two galaxies,
but we might be seeing the environmental dependence of galaxy formation
in the sense that the galaxy formation is accelerated in high density region.

\begin{figure}
\centering\includegraphics[width=80mm]{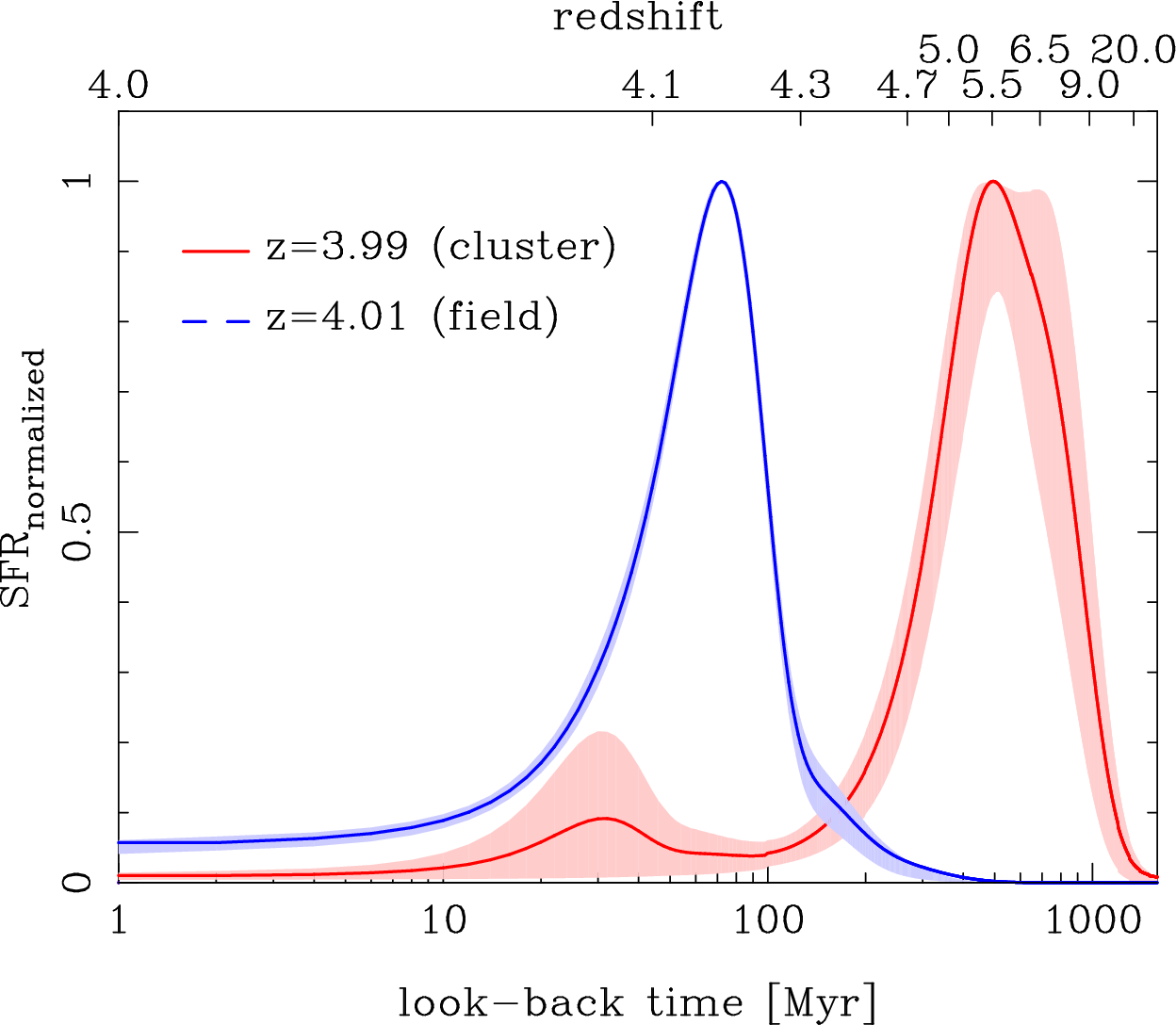}
\caption{
  Star formation histories of the $z=3.99$ (red) and $z=4.01$ (blue) galaxies.  The latter object is
  from \citet{tanaka19}.  The bottom axis shows the look-back time from each galaxy. The top ticks
  show the corresponding redshift from $z=4$ (i.e., mean observed epoch).
  We here use the mean redshift of the two galaxies just for clarify, and the small redshift difference
  of $\Delta z=\pm0.01$ from the spectroscopic redshifts does not strongly affect our discussions here.
  The vertical axis is in arbitrary unit and each SFH is normalized to unity at the peak.
  The shades show the statistical uncertainties.
}
\label{fig:age_distrib}
\end{figure}

\section{Joint Photometric and Spectroscopic Fits}
\label{sec:spectro-photometric_fit}

We turn our attention to the faint quiescent galaxies in the over-density region.
While direct redshift measurements from the spectra were not possible for them due to their
low-S/N (the median S/N per pixel is 0.8),
the spectra trace the continuum shape around the 4000\AA/Balmer break well.
We here make an attempt to constrain their redshifts by using both the spectra and
photometry. As our spectra are of low-S/N and there are residuals of the night sky lines,
we first bin the spectra by 100\AA\ by clipping the top and bottom 10\% of the flux
distribution in each bin. The clipping is intended to exclude the residual of the sky lines,
which is non-Gaussian and often contaminates the mean flux. We then feed the binned spectra
and photometry to a photometric redshift code \citep{tanaka15} to constrain their redshifts.

As a proof of concept, the technique is first applied to the confirmed $z=3.99$ and $z=4.01$ galaxies.
This ``spectro-photometric'' redshift technique successfully reproduces the original spectroscopic
redshifts of these two galaxies as shown in Fig.~7.
The redshift probability distribution functions are sharply peaked at the spectroscopic redshifts
with very small uncertainties.  These strong redshift constraints come from the continuum shape
traced by the MOSFIRE spectra; the spectra cover the break feature and the location of the break
gives a strong redshift constraint. We perform the same exercise for the spectroscopic objects
presented in \citet{ito23a} and confirm that we can successfully reproduce the spectroscopic
redshifts.

\begin{figure*}
\centering
\includegraphics[width=80mm]{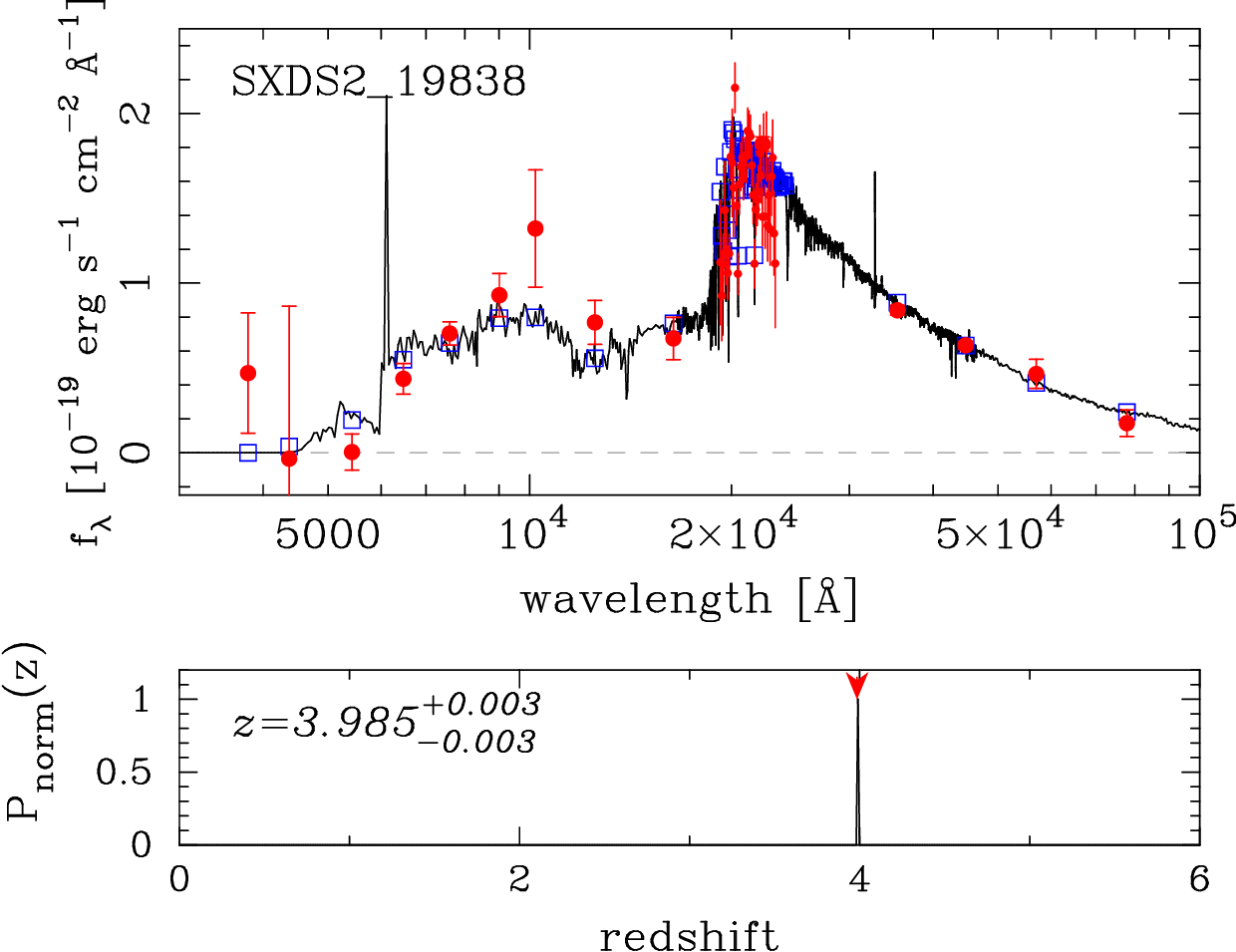}\hspace{0.5cm}
\includegraphics[width=80mm]{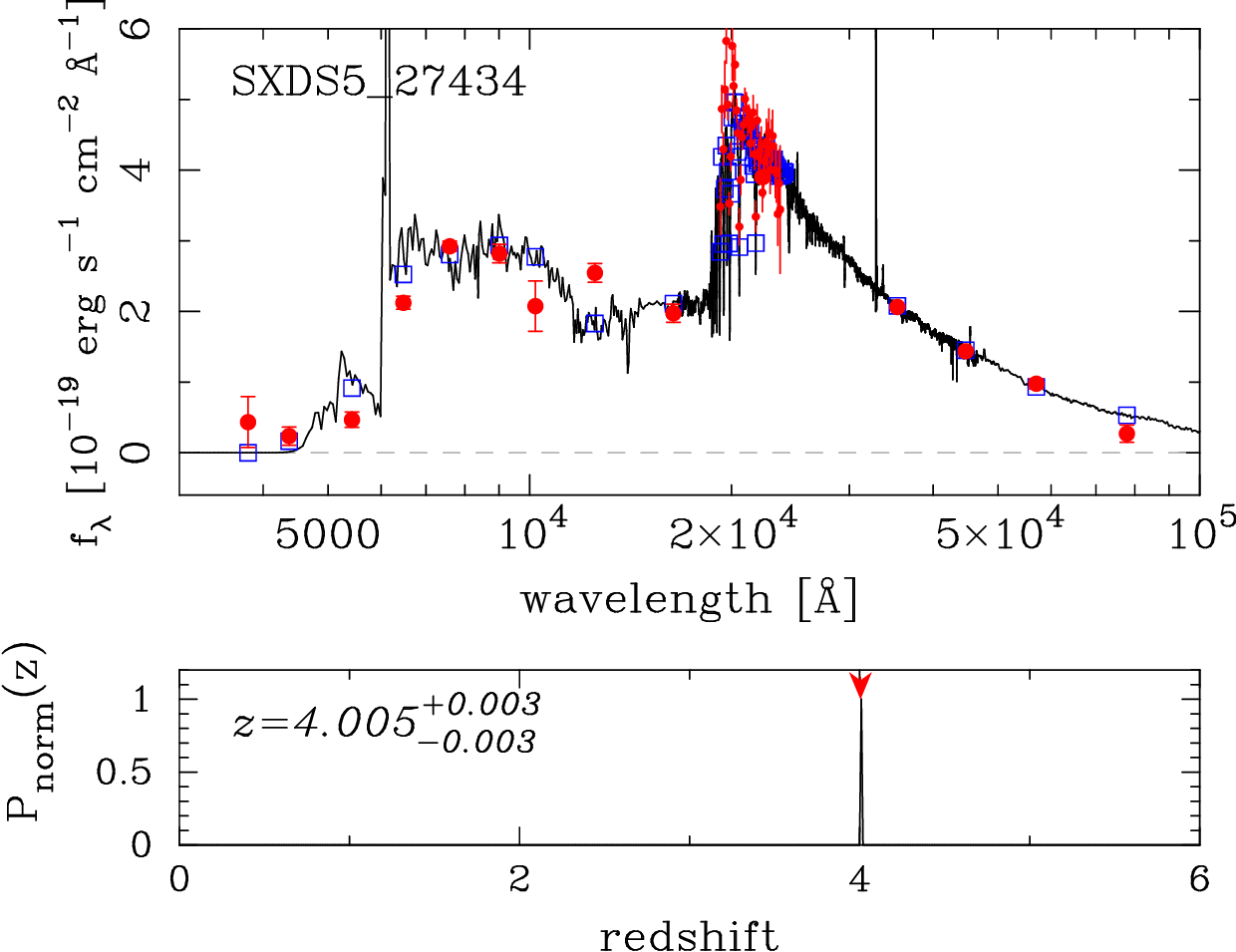}
\caption{
  Spectro-photometric fitting of the $z=3.99$ (left) and $z=4.01$ (right) galaxies.
  The top panel shows the observed SED.
  The large points are the broad-band photometry and the small points around $\lambda\sim2\mu\rm m$ are
  the binned MOSFIRE spectrum.
  The black spectrum is the best-fitting model template and the blue squares are the model fluxes.
  The bottom panel shows the redshift probability distribution function.
  It is sharply peaked at the spectroscopic redshift for both objects.
}
\label{fig:sed1}
\end{figure*}

Motivated by the excellent redshift recovery, we apply the technique to the faint quiescent galaxy
candidates for which no secure redshifts could be measured from MOSFIRE.
Fig.~8 shows the result; all of them show a sharp redshift spike at $z=4$.
There is one object with a significantly bimodal $P(z)$ distribution (SXDS2-16609), but the 2nd peak is consistent with $z=4$.
Due to the location of the object in the MOSFIRE mask, we could not cover the blue part of the $K$-band
for this object. That likely resulted in the wider $P(z)$.
In any case, it is very likely that most of these faint objects are located at $z=4$.
The inferred redshifts of the faint quiescent galaxies re summarized in Table 1.

\begin{figure*}
\centering
\includegraphics[width=80mm]{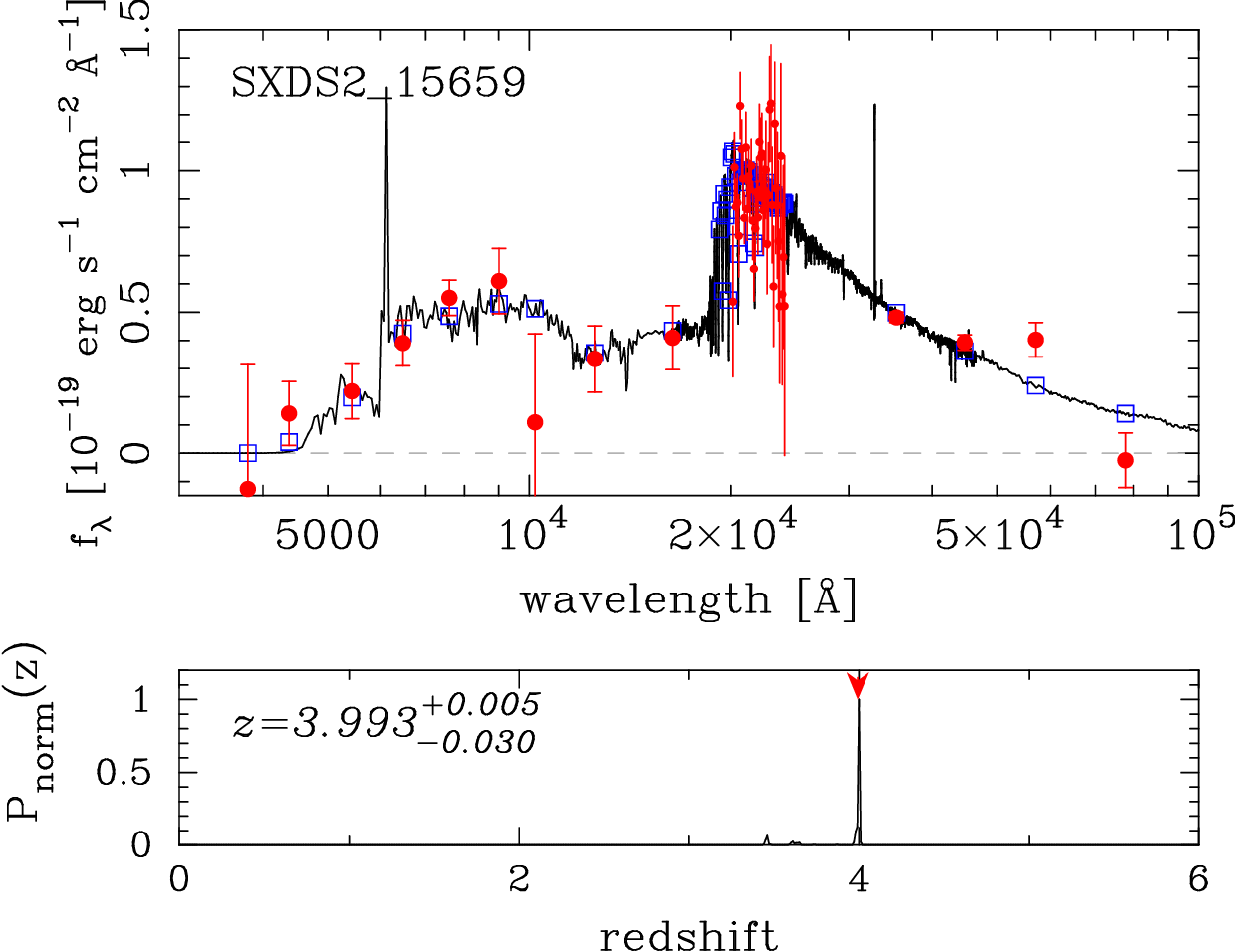}\hspace{0.5cm}
\includegraphics[width=80mm]{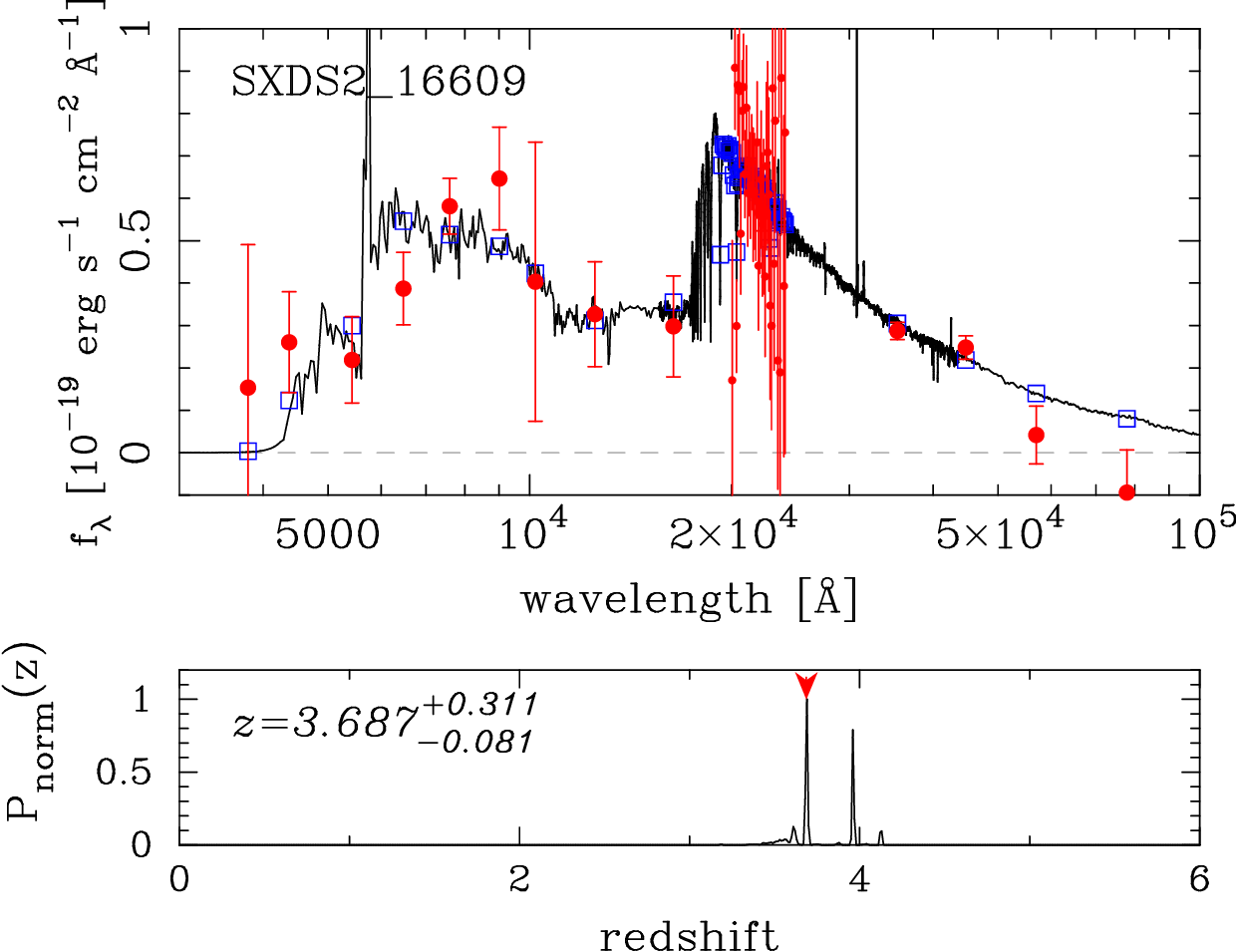}\\\vspace{0.5cm}
\includegraphics[width=80mm]{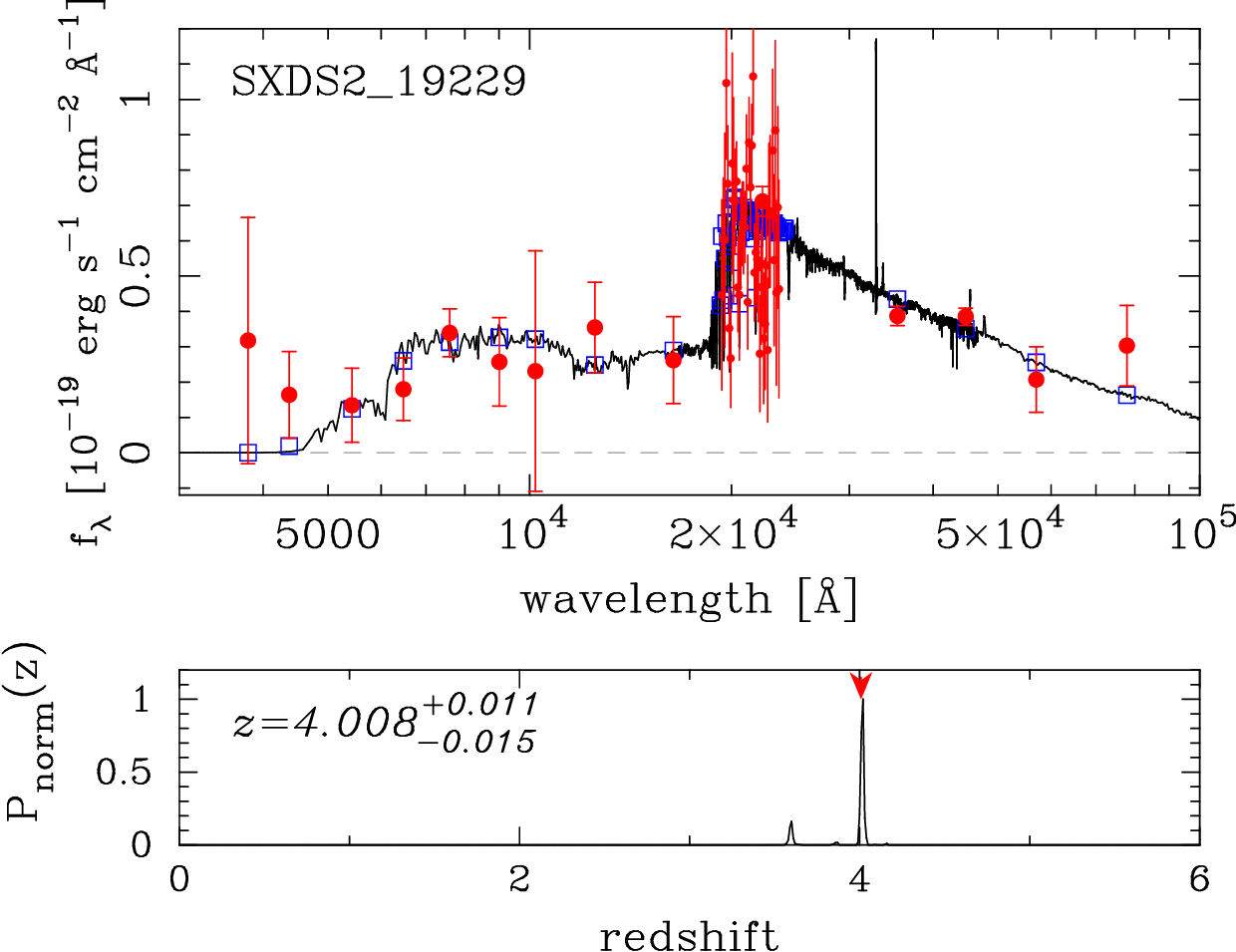}\hspace{0.5cm}
\includegraphics[width=80mm]{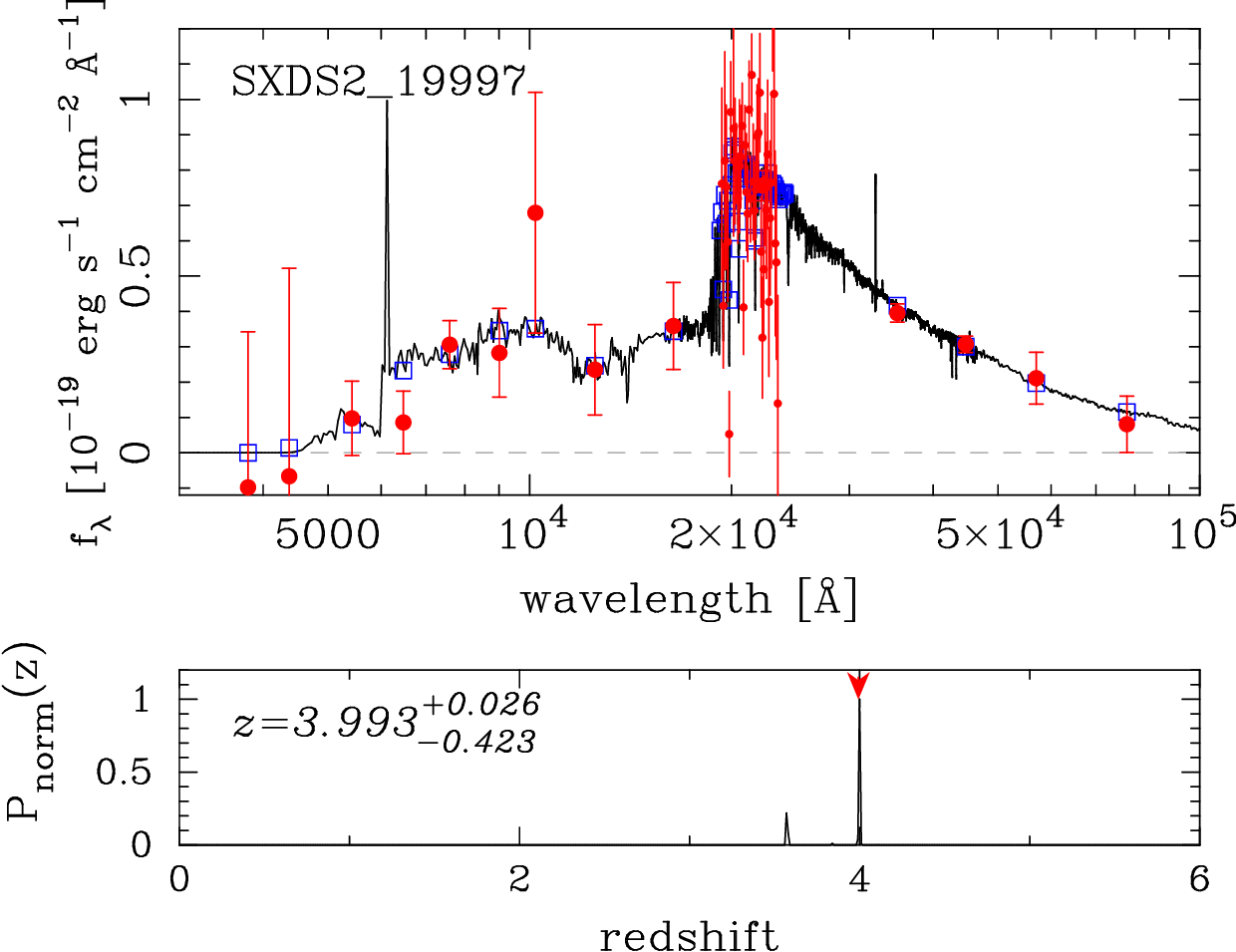}
\caption{
  Spectro-photopmetric fits to the faint quiescent galaxy candidates.  The meanings of the symbols are
  the same as in Fig.~7\\
}
\label{fig:sed2}
\end{figure*}

The spectro-photometric fits also give fairly robust estimates of physical properties of galaxies such
as stellar mass and SFR (Table~2).  Fig.~9 plots SFR against stellar mass.
The confirmed quiescent galaxies at $z=4$ are all below the star formation main sequence,
and they tend to be massive galaxies.
There are less quiescent galaxies and more star-forming galaxies at lower mass.
We are complete to $\rm M_*\sim3\times10^{10}M_\odot$
and the decreasing quiescent galaxies towards lower mass is likely a real trend.
We will elaborate on this point in Section 5.2.

\begin{figure}
\centering
\includegraphics[width=80mm]{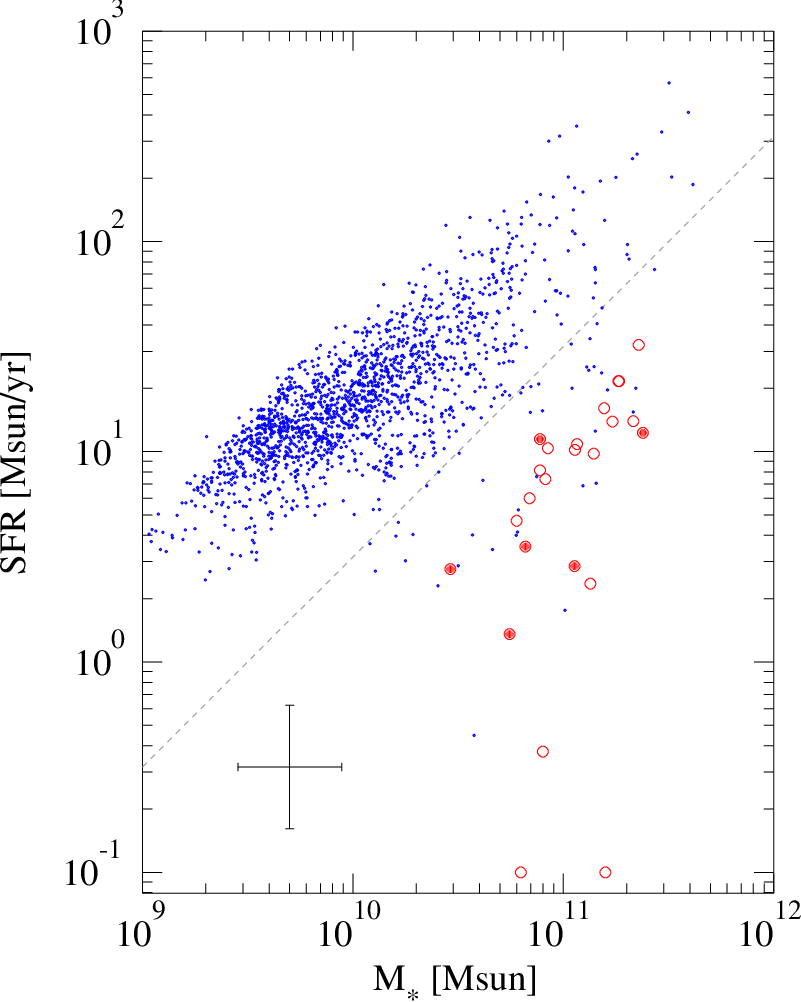}
\caption{
  SFR plotted against stellar mass for galaxies at $z\sim4$.
  The blue dots and red open circles are star-forming and quiescent galaxies at
  $3.7<z_{phot}<4.3$, respectively. 
  The error bars at the bottom-left corner indicate the typical fractional uncertainties
  including uncertainties in $z_{phot}$.
  The filled red circles are the quiescent galaxies
  at $z=4$ with secure spectroscopic/spectro-photometric redshifts.
  Recall that the quiescent galaxies are defined as those with a $1\sigma$ upper limit of
  sSFR being below $10^{-9.5}\rm\ yr^{-1}$ as indicated by the dashed line.
  Galaxies below SFR$<0.1\rm M_\odot\ yr^{-1}$ are all shown at SFR$=0.1\rm M_\odot\ yr^{-1}$.
}
\label{fig:sfr_smass}
\end{figure}

\section{Large-Scale Structure at $z=4$}
\label{sec:large_scale_structure}

\subsection{A cluster of quiescent galaxies}
\label{subsec:cluster_of_quiescent_galaxies}

Fig.~10 shows a zoom-in view of the over-density region.
The quiescent galaxies are located in a small region on the sky ($\sim2.5$ arcmin radius, which is $\sim1$ physical Mpc).
This is the densest region of quiescent galaxies discovered to date at such a high redshift;
the projected density of quiescent galaxies is $\sim1.5\rm~ Mpc^{-2}$.
We refer to it as cluster for now for simplicity.
Interestingly, the distribution of spectroscopically confirmed star-forming galaxies from VANDELS DR4 \citep{garilli21}
with secure redshift flags reveals a large-scale structure extending in the N-S direction.
The $z=4$ cluster is clearly  part of the structure, which adds further confidence.
The spectroscopically confirmed $z=4.01$ galaxy is located in the northern extension of the structure, and
the structure probably extends outside of the VANDELS coverage.
As mentioned earlier, the close redshifts of the spectroscopic objects ($z=3.99$ and $z=4.01$) are not
coincidence; they likely belong to the same structure.

\begin{figure*}
\centering
\includegraphics[width=160mm]{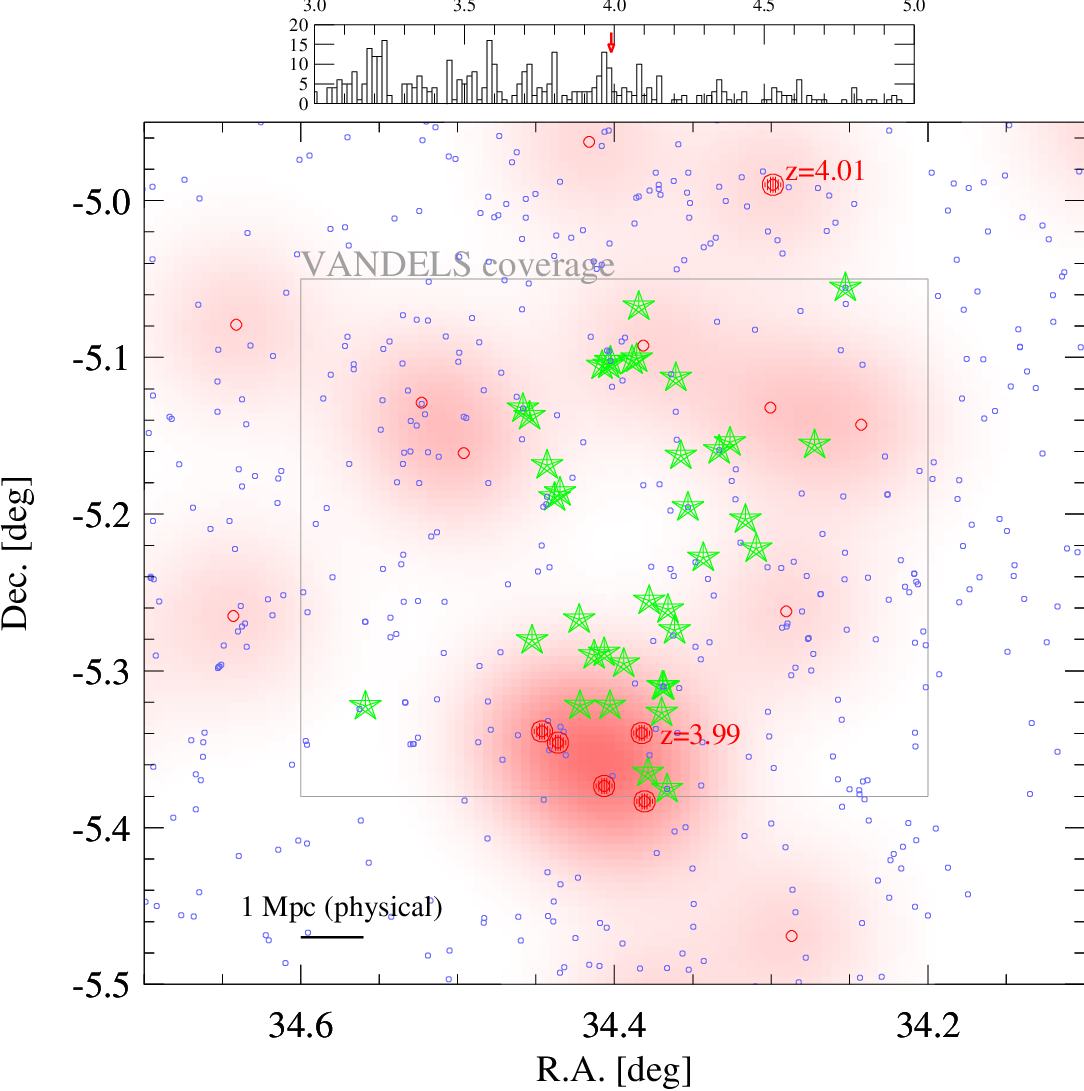}
\caption{
  Blow-up of the cluster region.
  The quiescent galaxies at $z\sim4$ are shown as the filled red points.
  Those with secure spectroscopic redshifts are indicated with the redshift labels.
  The green stars are spectroscopically confirmed $3.95<z_{spec}<4.05$ galaxies from VANDELS,
  and gray box indicates the region surveyed by VANDELS.
  The meaning of the other symbols are the same as in Fig.~2.
  The small panel on the top shows the redshift distribution of the VANDELS galaxies.
  The arrow indicates $z=3.99$. There is a redshift spike at $z\sim3.98$. The distribution
  of the green stars in the main panel remains virtually the same if we adopt a narrower
  redshift slice of, e.g., $3.95<z<4.01$.
}
\label{fig:z4_distrib_blowup}
\end{figure*}

A large number of (proto-)clusters at $z\gtrsim3$ have been reported in the literature, but all of
them used star-forming galaxies as a tracer \citep{miller18,oteo18,toshikawa18}.  The current frontier of bona-fide clusters with red sequence
is $z\sim2$ \citep{gobat11,spitler12}, and recent work has just started to imply possible higher redshift clusters \citep{mcconachie22,ito23a}.
Our cluster is the most distant cluster of quiescent galaxies discovered to date 
and is an ideal laboratory to address the role of environment at the highest redshift ever probed.

A ubiquitous feature of low-redshift clusters is the red sequence on a color-magnitude diagram \citep{bower92} formed
by quiescent galaxies. We present the color-magnitude diagram in Fig. 11. Quiescent galaxies
tend to be red as expected and there is a relatively large color scatter among quiescent galaxies.
However, the quiescent galaxies in the cluster shown by the filled red circles seem to exhibit
a smaller color scatter and form an early red sequence.
Most of the quiescent members do not exhibit large dust extinction (Table 2)
and it is likely that their red colors are due to their ages.
As we are very close to the primary formation epochs of these galaxies, the observed color is sensitive
to a small difference in their formation redshifts.
The small color scatter implies that the cluster galaxies form (and quench) about the same time.
They distribute over a 1 Mpc scale and they may not form a fully virialized system yet, but
our observation here seems to suggest that the red sequence first appears during the initial collapse of a massive halo,
accompanied by the roughly simultaneous quenching event in each halo.

This picture is different from the pre-processing
scenario suggested at much lower redshifts, in which galaxies become quiescent in low-mass groups before they fall
into more massive halos (e.g., \citealt{kodama01,fujita04}). Environmental effects are likely stronger at lower redshifts with more abundant intracluster
medium, but any effects due to the intracluster medium are not important in our case because our cluster is
likely in a pre-collapse phase (see below). Possible quenching processes may include gas-exhaustion due to
starburst triggered by galaxy-galaxy interactions. Interactions may enhance AGN activities, which may then trigger
AGN feedback to further suppress star formation. Recent work by \citet{kakimoto23} reported on a massive galaxy
being quenched in a very compact group at $z=4.5$ and suggested a possible role of mergers for quenching at
high redshifts. Their galaxy indeed has a very close companion, and it
is interesting to note that this is not the only case where a massive quiescent galaxy accompanies a close companion
(e.g., \citealt{schreiber18b}).
A close companion may indicate a potential role of very small-scale environment for quenching.

Star-forming galaxies, on the other hand, are separated from the red sequence and form a cloud of blue galaxies,
which indicates that the bimodality of galaxy populations is already in place as early as $z = 4$.
Overall, the segregation of galaxy properties that we observe in the local Universe has already emerged at this epoch.

\begin{figure}
\centering
\includegraphics[width=80mm]{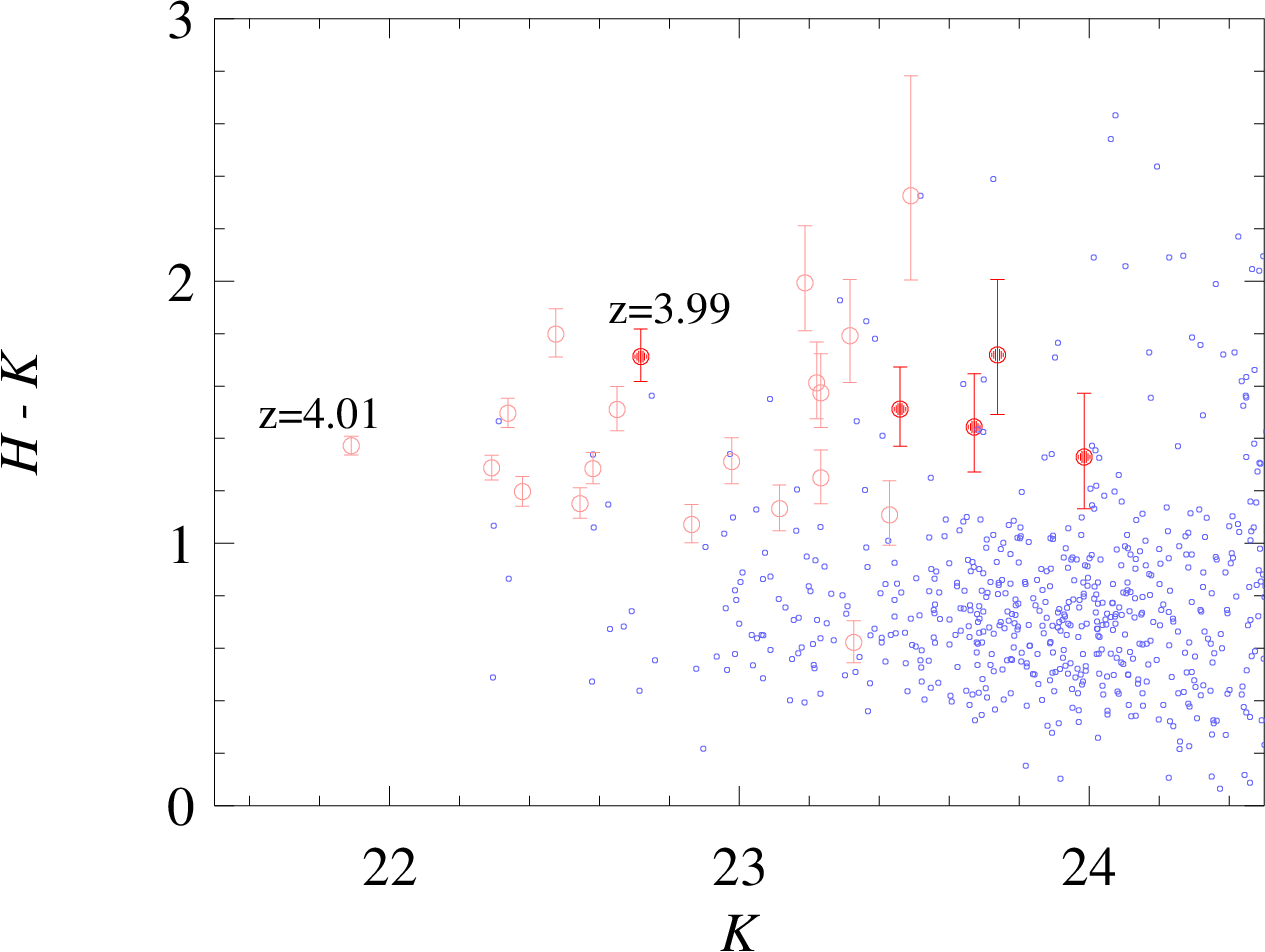}
\caption{
  Observed $H-K$ color plotted against $K$-band magnitude.
  The meanings of the symbols are the same as in Fig.~9,
  except that the filled red circles show the quiescent galaxies in the over-density region only
  (i.e., $z=4.01$ galaxy is excluded).
  Objects with secure redshifts are indicated.
}
\label{fig:cmd}
\end{figure}

\subsection{The Butcher-Oemler effect at $z=4$}
\label{subsec:butcher_oemler_effect}

We further discuss the environmental dependence of galaxy SFRs by extending the Butcher-Oemler effect
\citep{butcher78} to $z=4$ for the first time. We first define the cluster region using a 1~Mpc aperture
that encloses the concentration of the quiescent galaxies and estimate the quiescent fraction.
Here, we include photometric
redshift selected star-forming galaxies within the aperture as blue members.
As in Fig.~2, we use galaxies at $3.7<z_{phot}<4.3$.
We apply the same selection to all galaxies outside the aperture to construct a field sample for comparison.
The fractions of quiescent galaxies are shown in Fig.~12.

We begin with the global trend using the entire (=cluster + field) sample.
The black points nicely show that the quiescent fraction is
a strong function of stellar mass in the sense that it increases towards higher mass.  The fraction
is as high as $\sim30\%$ at $\rm M\gtrsim10^{11}\ M_\odot$. On the other hand, it is very low at
low mass and low-mass galaxies are predominantly star-forming galaxies. This trend is consistent
with \citet{weaver22}. In other words, quiescent galaxies are almost exclusively massive galaxies.

Turning to the cluster region, we do not have sufficient statistics to discuss the quiescent
fraction as a function of stellar mass, and we instead estimate the quiescent fraction using
all the cluster members. We take the mean stellar mass here and show the quiescent fraction in Fig.12
as the filled red circle. The quiescent fraction is about 35\%.
For comparison, we do the same for the field galaxies and show their quiescent fraction as
the blue circle. The field quiescent fraction is only 1.5\% and is significantly lower than
that of the cluster. While we have only a single cluster, our finding here indicates that
a cluster environment exhibits a higher quiescent fraction from an early phase of its formation.
In other words, the Butcher-Oemler effect might be already there at $z=4$.

A larger cluster sample is clearly needed to confirm the trend, but it will be a significant
challenge to identify a cluster like ours as one needs to survey a large volume and then push
the current spectroscopic facilities to the limit to obtain secure redshifts.  Nonetheless,
it is worth the effort to search for more clusters at higher redshifts to peer deeper into
the cluster formation.

\begin{figure}
\centering
\includegraphics[width=80mm]{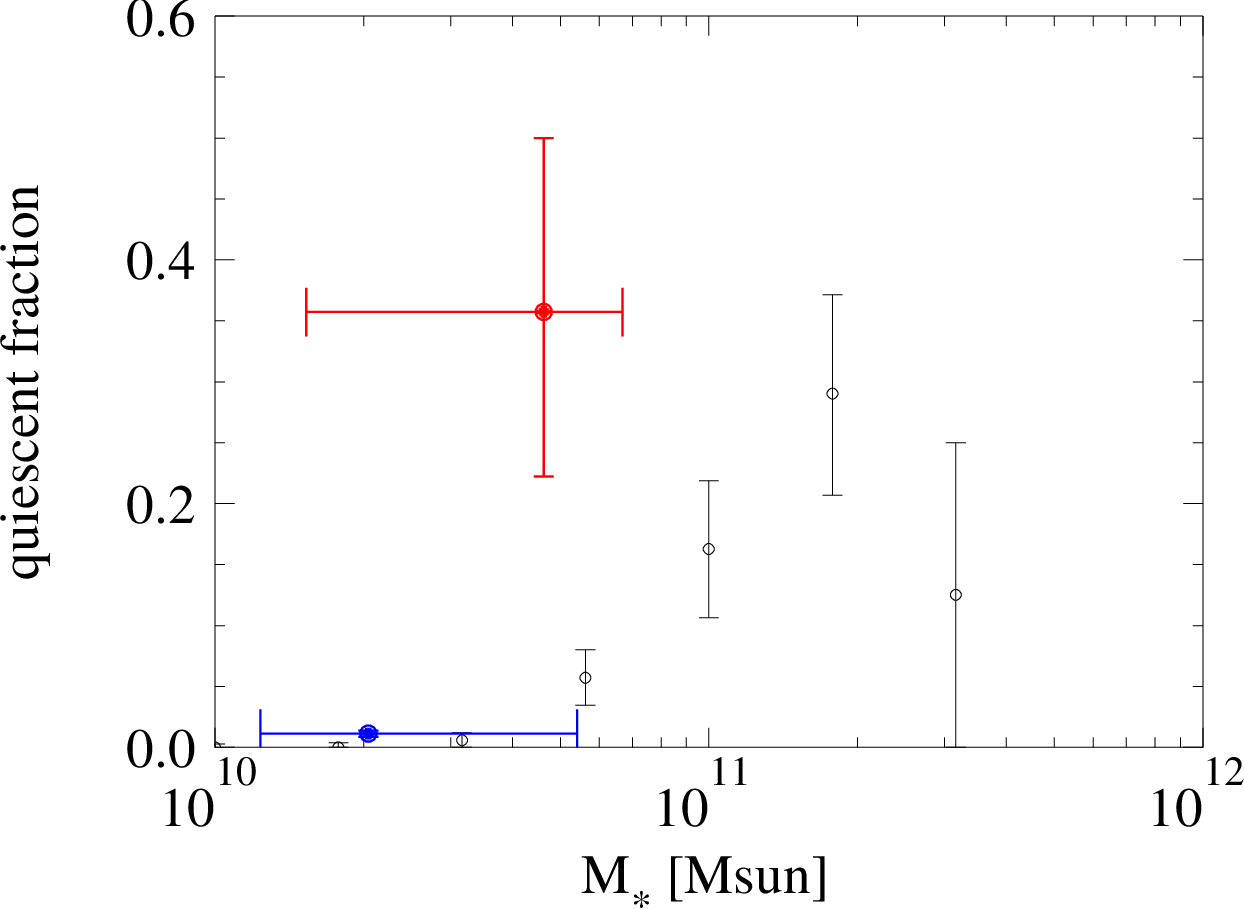}
\caption{
  Fraction of quiescent galaxies as a function of stellar mass based on galaxies $3.7<z_{phot}<4.3$.
  The black open circles are for the combined field and cluster sample to see the global trend
  and are plotted as a function of stellar mass.
  The red and blue filled circles are the average quiescent fractions of the cluster and field regions, respectively.
  These circles are plotted at the mean stellar mass.
  The vertical error bars are the Poisson errors in the quiescent fraction, while the horizontal error bards
  show the 68\% interval of the stellar mass distribution.
}
\label{fig:qfrac}
\end{figure}

\begin{deluxetable*}{lcccc}
  \tablecaption{
    Properties of the quiescent galaxies
  }
  \tablehead{
    \colhead{ID} & \colhead{$\rm M_* (10^{10}M_\odot)$} & \colhead{$\rm M_{halo} (10^{12}M_\odot)$} & \colhead{$\rm SFR (M_\odot\ yr^{-1})$} & \colhead{$\tau_V$}
  }
  \startdata
  SXDS2\_19838 & $11.31^{+0.09}_{-0.61}$ & $3.1^{+6.6}_{-2.0}$  & $2.9^{+0.0}_{-0.1}$  & $0.1^{+0.0}_{-0.1}$\\
  SXDS2\_15659 & $6.60^{+0.78}_{-0.72}$  & $2.3^{+4.6}_{-1.4}$  & $3.5^{+3.3}_{-0.5}$  & $0.3^{+0.2}_{-0.1}$\\
  SXDS2\_16609 & $2.91^{+0.56}_{-0.42}$  & $1.5^{+2.7}_{-1.0}$  & $2.8^{+0.8}_{-1.4}$  & $0.1^{+0.2}_{-0.1}$\\
  SXDS2\_19229 & $7.76^{+0.83}_{-1.13}$  & $2.6^{+5.7}_{-1.7}$  & $11.5^{+7.9}_{-6.0}$ & $0.9^{+0.1}_{-0.3}$\\
  SXDS2\_19997 & $5.55^{+0.47}_{-1.06}$  & $2.1^{+3.6}_{-1.4}$  & $1.4^{+0.2}_{-0.7}$  & $0.1^{+0.1}_{-0.1}$\\
  \hline
  SXDS5\_27434 & $23.99^{+0.10}_{-0.09}$ & $5.3^{+12.3}_{-3.7}$ & $12.3^{+0.1}_{-0.1}$ & $0.0^{+0.1}_{-0.0}$\\
  \enddata
  \tablecomments{
    The physical properties are based on the joint photometric and spectroscopic fits.
    The third column is the inferred halo mass (see text for details).
    The last column ($\tau_V$) is the dust extinction optical depth in the $V$-band.
  }
  \label{tab:props}
\end{deluxetable*}


\section{Summary and Discussion}
\label{sec:discussion}

We have presented the spectroscopic confirmation of a massive quiescent
galaxy at $z=3.99$ as well as the discovery of a significant concentration of
quiescent galaxies at $z=4$.  Based on the deep spectroscopic observation
with MOSFIRE, we have identified multiple Balmer absorption lines of
the $z=3.99$ quiescent galaxy. Its stellar velocity dispersion is $\sigma=305\pm103\rm\ km\ s^{-1}$ and is slightly
higher than lower redshift counterparts with similar stellar masses, indicating
a mild redshift evolution. Its physical
size measured from the deep imaging data reveals its compact size and dense
stellar density, consistent with other quiescent galaxies observed at
similar redshifts in the literature.

Interestingly, the galaxy is located in an over-density region of quiescent galaxies
at similar redshifts.  While we could not directly measure their spectroscopic
redshifts from their spectra, the high-accuracy spectro-photometric redshifts
strongly suggests that they are at the same redshifts as the $z=3.99$ galaxy.
All these galaxies exhibit a pronounced Balmer break, indicative of recent starburst
activity followed by rapid quenching.  Their UV emission is relatively weak and their SFRs inferred from their
overall SEDs indeed suggest that they are quiescent.  The concentration is embedded in a large-scale filament
traced by the VANDELS spectroscopic galaxies, which adds further credits to the physical association.  This is the first discovery of
a concentration of quiescent galaxies at this high redshift and it serves as an
unique laboratory to investigate the role of environment at $z=4$ for the first time.

The cluster galaxies form the early red sequence, suggesting that these galaxies form
and quench at similar times, and the red sequence emerges
during the first collapse of a cluster.  In addition, the quiescent galaxy fraction is
significantly higher than the field average.  This may be expected because we define the over-density
region around the quiescent galaxies, but we have extended the Butcher-Oemler effect to
z=4 for the first time. The fact that the collapsing cluster exhibits a higher quiescent fraction suggests that we may be witnessing the
environmental dependence of galaxy formation such that galaxies in higher density regions form earlier than the field galaxies.
We aim to address this point further in our future paper.

Given the significant over-density, we expect that
the cluster region will collapse to form the core of a massive cluster at lower redshifts.
We make an attempt to
roughly estimate the future halo mass of the system when they collapse.
Using the halo occupation distribution modeling from \citet{shuntov22}, we infer
the halo mass of each quiescent galaxy from its stellar mass as summarized in Table 2.
The individual halo mass ranges from 1 to $3 \times 10^{12}\rm~ M_\odot$. If we assume that
they all collapse to form a single halo, the total mass exceeds $10^{13}\rm~ M_\odot$, i.e., group-sized halo.
We consider only the quiescent members, but there are star-forming galaxies with consistent photo-$z$'s,
and thus the total halo mass here is a lower limit. We find that the virial radii of the galaxies derived from their halo
mass do not overlap with each other, suggesting that they are not yet in the same halo at the time of
the observation.

In order to further characterize the properties of the cluster, we have searched for similar systems in
the Illustris TNG-300 simulation at $z=4.01$ (snapshot 21; \citealt{nelson19}).
To be specific, we identify quiescent galaxies with $\rm M_*>10^{10}\rm M_\odot$ and $\log \rm sSFR<10^{-9.5}\ M_\odot\ yr^{-1}$
in the simulation and examine their spatial distribution.
Stellar mass and SFR are measured within twice the stellar halo-mass radius, and SFR is
averaged over 10~Myr.
Interestingly, there is no over-density region of the quiescent galaxies in TNG-300; there are only 11 massive
quiescent galaxies within the 300 comoving Mpc simulation box and the closest separation between
them is as large as 8 physical Mpc (3d distance).  In contrast, we have 5 quiescent galaxies within
1 physical Mpc radius (projected distance).
While the 3d and projected distances are different, it is clear that the observed cluster is a much denser system.
We note that we perform the same analysis but considering more massive galaxies with $>5\times10^{11}\rm M_\odot$ only
and confirmed there is still no counterpart in Illustris TNG-300.

We do not know yet if this is a failure of the simulation in reproducing quiescent galaxies at $z\sim4$;
the number density of massive quiescent galaxies at $3\lesssim z\lesssim4$ is in broad agreement
between observation and simulations (\citealt{valentino23} and references therein), although
it depends strongly on how quiescent galaxies are defined.
It may be that our system is so rare that the simulation box is too small to reproduce it.
In any case, massive halos at high redshifts have to be rare and it is crucial to increase
the cluster sample to perform further comparisons with simulations.

We expect that our spectra of the faint cluster members will be able to strongly constrain their star formation
histories as the spectra cover the Balmer/4000$\rm\AA$ break for most of them. All of the members show
a sign of suppressed star formation as discussed earlier and it would be of interest to perform extensive
SED fitting to infer their primary formation epoch and quenching timescale and compare them with those of
the field galaxies from the literature. Also, we can measure the physical sizes of the members and see
whether there is environmental dependence of galaxy sizes at these redshifts. The environmental dependence of
physical sizes of galaxies at low redshifts is somewhat controversial in the literature (see Table 1 of \citealt{yoon17},
and Table A1 of \citealt{matharu19}),
and it would be interesting to look at the $z=4$ system. We hope to address these points in a forthcoming paper.
Also, detailed imaging/spectroscopic follow-up observations with JWST are an interesting avenue as they
deliver secure spectroscopic redshifts, more accurate physical sizes, etc. ALMA observations to search for
dust-obscured galaxies that went unnoticed in our work will be interesting, too. This work can be extended in
multiple ways, and joint analyses will hopefully lead us to deeper insights into the early cluster formation and
the quenching physics in the early Universe.

\begin{acknowledgments}

This work is supported by JSPS KAKENHI Grant Numbers JP23740144, 22KJ0730, JP15K17617.
The Cosmic Dawn Center (DAWN) is funded by the Danish National Research Foundation (DNRF140).

Some of the data presented herein were obtained at the W. M. Keck Observatory,
which is operated as a scientific partnership among the California Institute of Technology,
the University of California and the National Aeronautics and Space Administration.
The Observatory was made possible by the generous financial support of the W. M. Keck Foundation.
The authors wish to recognize and acknowledge the very significant cultural role and reverence
that the summit of Maunakea has always had within the indigenous Hawaiian community.
We are most fortunate to have the opportunity to conduct observations from this mountain.

The HSC collaboration includes the astronomical communities of Japan and Taiwan, and
Princeton University. The HSC instrumentation and software were developed by the National
Astronomical Observatory of Japan (NAOJ), the Kavli Institute for the Physics and Mathematics of
the Universe (Kavli IPMU), the University of Tokyo, the High Energy Accelerator Research
Organization (KEK), the Academia Sinica Institute for Astronomy and Astrophysics in Taiwan (ASIAA),
and Princeton University. Funding was contributed by the FIRST program from Japanese Cabinet Office,
the Ministry of Education, Culture, Sports, Science and Technology, the Japan Society for
the Promotion of Science, Japan Science and Technology Agency, the Toray Science Foundation,
NAOJ, Kavli IPMU, KEK, ASIAA, and Princeton University.  This paper makes use of software
developed for the Large Synoptic Survey Telescope. We thank the LSST Project for making their
code available as free software at  http://dm.lsst.org. This paper is based in part on data
collected at the Subaru Telescope and retrieved from the HSC data archive system, which is
operated by Subaru Telescope and Astronomy Data Center at NAOJ. Data analysis was in part
carried out with the cooperation of Center for Computational Astrophysics, NAOJ.

VANDELS is Based on data products from observations made with ESO Telescopes at the La Silla
Paranal Observatory under program ID 194.A-2003(E-K).
\end{acknowledgments}

\facilities{Keck(MOSFIRE)}


\appendix
\section{Data Reduction}

In this Appendix, we detail our data reduction procedure. 
Each A-B pair was flat-fielded and wavelength-calibrated by using common flat and wavelength
reference frames. The background was removed by subtracting the frame at the B dither position
from that at the A dither position followed by subtracting residuals \citep{kelson03}.
Finally, all A-B pairs were rectified to the same wavelength solution, while coaddition was
done for each A-B pair separately by applying the dither offset at this stage. The slit mask
alignment procedure was carried out every $\sim 2$ hours, but there were still noticeable slit
drifts in the spatial direction between the alignment procedures \citep{kriek15,larson22}.
We placed a bright point source (SXDS2\_19063, a quasar) to monitor the drift and the seeing
variation. The drifts up to $\sim2$ pixels were corrected to locate objects in a common spatial frame.

We observed an A0V star (HIP 111996) for telluric correction twice per night with the {\tt long2pos}
slit mask which is designed to take an AB dither at 2 positions (Pos A and Pos C), which enables
us to cover the entire wavelength range of $K$-band. We reduced spectra with the DRP at each position
separately and extracted one-dimensional spectra by applying the optimal extraction method \citep{horne86}
assuming a Gaussian spatial profile. The extracted spectra were normalized at $2.1\mu m$ and then combined.
We used Pos A spectra at $\lambda > 2.3 \mu m$, while spectra at Pos C were used for the rest of
the wavelength range. This is because we found that the S/N is higher at Pos C most likely due to
a misalignment of the slit geometry and preset of the telescope offsets. The telluric correction vectors
were created by dividing the telluric standard spectra by a Vega template spectrum \citep{kurucz93}
after convolving to the MOSFIRE resolution of R=3600. The telluric correction vector was applied for
each two-dimensional spectrum closest in time to the observation time of HIP 111996.

The coaddition of two-dimensional A-B stacks was carried out by taking a weighted mean using the peak
counts measured for the point source used for the drift correction. Two among 159 A-B stacks were not
used since they showed too low S/N to robustly derive the traces of the source due to cloudy conditions.
We also applied 3-sigma clipping to remove bad pixels and cosmic rays.  Error spectra were derived by
propagating error spectra associated with each A-B stack. The total integration time of the coadded
spectra used in the following analysis is 15.7 hours.

Absolute flux calibration was made by using the point source common for all science exposures.
We used an optimally extracted one-dimensional spectrum for the object and derived a synthetic
$Ks$-band magnitude by convolving it with the VISTA Ks-band filter response curve for each A-B stack.
We also corrected the slit loss by comparing the slit width and a Gaussian FWHM measured along
the spatial direction. Then, we applied normalization factors by comparing the synthetic magnitudes
with the observed VISTA Ks-band total magnitude \citep{mehta18} to scale the spectra to the absolute fluxes.
The spectra fully calibrated in this way are used in the main body of the paper.

\bibliography{references}{}
\bibliographystyle{aasjournal}

\end{document}